\documentclass[12pt]{report}
\usepackage{amsmath,amssymb,bbold,hyperref,color,braket,xcolor,graphicx,float}
\DeclareMathOperator{\csch}{csch}
\newcommand{\RNum}[1]{\uppercase\expandafter{\romannumeral #1\relax}}
\usepackage{simpler-wick}
\usepackage[autostyle]{csquotes}
\usepackage[titles]{tocloft}

\hypersetup{pdfborder=0 0 0}
\newcommand{\squeezeup}{\vspace{-6mm}}
\newmuskip\pFqmuskip

\newcommand*\pFq[6][8]{%
  \begingroup 
  \pFqmuskip=#1mu\relax
  \mathchardef\normalcomma=\mathcode`,
  \mathcode`\,=\string"8000
  \begingroup\lccode`\~=`\,
  \lowercase{\endgroup\let~}\pFqcomma
  {}_{#2}F_{#3}{\left[\genfrac..{0pt}{}{#4}{#5};#6\right]}%
  \endgroup
}
\newcommand{\pFqcomma}{{\normalcomma}\mskip\pFqmuskip}
\begin{document}
\begin{titlepage}
\begin{center}
\vspace{10mm}
{\LARGE \textbf{Study of Information Restoring Effects in Virasoro Blocks}}
\vspace{12mm}

\renewcommand\thefootnote{\mbox{$\fnsymbol{footnote}$}}
\large{Shivrat Sachdeva}${}^{1}$\footnote{shivrats@imsc.res.in}

\vspace{10mm}

${}^1${\small \sl The Institute of Mathematical Sciences} \\
{\small \sl 4th Cross St, CIT Campus, Tharamani, Chennai 600113 ,India}

\end{center}
\vspace{12mm}
\noindent
It is well-known that in the large central charge \(c\) limit, the Virasoro blocks in CFT\(_2\) suffer from information loss when studying the BTZ black holes in the dual AdS\(_3\) theory. Recent studies have proved that non-perturbative \(e^{-c}\) corrections resolves information loss. Here, we study universal features of the spectrum of heavy states from general properties of finite \(c\) Virasoro conformal blocks. Furthermore, we investigate the behavior of correlators beyond “Page” time \(t \sim S_{BH}\), entropy of the black hole. In the Lorentzian regime, we wish to better understand the transition from an exponential decay behaviour of Virasoro blocks in the heavy-light limit to a power-law decay at late times.
\end{titlepage}
\setcounter{footnote}{0}
\renewcommand\thefootnote{\mbox{\arabic{footnote}}}
\chapter*{Declaration}
This project which was done as a part of the Master's in Science (Physics) at Harish Chandra Research Institute is an account for the ongoing research which started from March 2024 at the Centre of High Energy Physics, Indian Institute of Science. The project is submitted in fulfillment of the requirements for the degree of Master's in Science (Physics). 
\vfill
\begin{flushright}
    \begin{tabular}{l}
   \vspace{25mm}
\begin{tabular}{@{}p{2.5in}@{}}
Submitted by: \\
Shivrat Sachdeva
\end{tabular}
    \end{tabular}
\end{flushright}
\chapter*{Acknowledgements}
I thank my advisor Chethan Krishnan of the \textit{Centre of High Energy Physics, Indian Institute of Science (IISc)} for providing me with an opportunity to work with him. I am extremely grateful to him for his support, for sharing his ideas with me and encouraging our long and frequent fruitful discussions. Furthermore, I am thankful to Dileep Jatkar of \textit{Harish Chandra Research Institute (HRI)} for sharing his immense knowledge, experience and nurturing my curiosity all throughout my time at \textit{HRI}. At last, I will forever be grateful for the unconditional support of my family and friends all throughout my journey.
\tableofcontents
\bigskip
\hrule
\addtolength{\parskip}{8pt}
\chapter{Introduction}
The theory of general relativity governs the behavior of gravitating objects ignoring the quantum gravity effects, and predicts the existence of a singularity in the black hole interior. These kind of singularities are generally a result of the large curvatures compared to the quantum scales. This is in contrast to our usual laboratory experiment where there is no overlap between quantum effects and gravity. However, in the black hole interior, explaining the strong interactions between gravity and quantum mechanics requires a presence of a quantum theory of gravity. Fortunately, due to Einstein's equations we know that larger the black hole, smaller is it's curvature. The larger the black hole is, the longer it will take for the gravitational effects to be strong enough that it requires a complete quantum treatment. This means that for a large part of the black hole interior, a semi-classical approximation is accurate enough. \\ \\ 
Unfortunately this innocuous and rudimentary result leads to one of the most famous and longstanding open problems in physics: the black hole information paradox \cite{PhysRevD.14.2460, Page_1993,Almheiri_2013,Mathur_2009}. This paradox arises from the conflict between two fundamental principles of modern physics: the first states that semiclassical gravity is a good approximation at energy scales where quantum effects are negligible; while the second states that the “unitarity” of quantum mechanics allows all quantum processes to be reversible. \\ \\ In 1975 Stephen Hawking showed that if semiclassical gravity is a valid approximation at the event horizon of a black hole, then black holes can evaporate. However, if the black hole evaporates it is impossible to reverse-engineer the information that went into a black hole before it evaporated. As the black hole evaporates it's emitted radiation must be thermal, and indistinguishable between any two evaporated black holes. This represents a catastrophic loss of information of an infalling observer. The black hole information paradox has been the beacon for new scientific ideas and a complete description of quantum gravity ever since it's discovery by Hawking. New ideas in string theory have provided us solid evidence that maybe information is not lost. However, the question how information can be conserved is unsolved. \\ \\ A few years ago, the work of AMPS \cite{Almheiri_2013} suggested an alternative to Susskind's \textit{black hole complementarity} (BHC) formalism \cite{Susskind_1993,Susskind_1994}. If we assume near the horizon instead of entangled ingoing bits, Hawking pairs are generated, then in that case the infalling observer must encounter high energy quanta at the horizon, which is commonly refereed to as the “black hole firewall”. \\ \\ Gauge/gravity duality provides convincing evidence that all information from the evaporated black hole is carried away its the Hawking radiation. Maldacena in his paper \cite{Maldacena_2003} proposed that an eternal black hole in AdS can be holographically described by considering two identical, non-interacting copies of the conformal field theory. Using this correspondence Maldacena investigated some aspects of the information loss paradox. Precisely speaking, he considered an AdS\(_{d+1}\) spacetime and its holographic dual conformal field theory CFT\(_d\), where the conformal field theory is defined on a cylinder \(R \times S^{d-1}\) and the cylinder acts like the boundary of the AdS spacetime. The information loss paradox becomes precise in AdS due to the existence of external black holes. The information loss argument in \cite{PhysRevD.14.2460, Page_1993, Mathur_2009,Almheiri_2013} states that all correlators with the infalling matter decay exponentially inside the black hole. Building on this idea, Maldacena considered a simple prototypical deformation of a perfectly thermal Schwarzschild AdS state. The two-point function for this new deformed thermal ensemble decays exponentially as \(e^{-t/\beta}\). This is a typical example where information loss in AdS is sharp. \\ \\ Different approaches have been taken to understand this paradox. One of which involves using the CFT tools to probe the AdS black hole physics. The Virasoro algebra is extremely constraining on the dynamics of CFTs in two dimension. It's common knowledge that studying quantum gravity in AdS\(_3\) requires us to analyze CFTs with large central charge \(c\). The Virasoro blocks have turned out to be extremely useful in understanding gravity in AdS\(_3\), and in fact BTZ black hole thermodynamics naturally emerge when considering heavy-light in the semi-classical large \(c\) limit of the Virasoro blocks. Apparently, information loss appears to be due to the behavior of the conformal blocks in this large \(c\) limit. The conformal blocks are also the bread and butter of the conformal bootstrap program. By knowing their explicit forms we can study 2d CFTs and 3d gravity using bootstrap. Although we have struggled understanding the information paradox, AdS/CFT has provided us with major leaps in better understanding the paradox.  \\ \\The semiclassical large central charge limit of the Virasoro blocks precisely matches the results which are directly computed from the AdS\(_3\) gravity side. In the semiclassical limit, the Virasoro blocks exhibit information loss in the form of ‘forbidden singularities’ and exponential decay at late Lorentzian times \cite{fitzpatrick2016conformalblockssemiclassicallimit,fitzpatrick2016informationlossads3cft2,Fitzpatrick_2011,Fitzpatrick_2014}. Information loss in BTZ black hole backgrounds pollutes the CFT correlators with contaminants which show up in the large \(c \propto \frac{R_{AdS}}{G_N}\) limit. This makes information loss in AdS\(_3\)/CFT\(_2\) extremely precise. A way to removing the forbidden singularities of Virasoro blocks is via non-perturbative effects in \(1/c\). The motivation for exploration in this direction was twofold. The information loss manifests as unitarity violation due to the forbidden singularities that arise in semiclassical correlators, with these singularities being present at each individual level of the Virasoro blocks.
Understanding large \(c\) saddles is critical in studying the quantum corrections involved in the restoration of unitarity.
\\ \\ Unfortunately, there is another big void in the quantum gravity description. We know classical black hole in the effective field theory (EFT) have smooth horizons with entropy \(S=\frac{A}{4G}\). However, in the bulk EFT, this entropy cannot be accounted for. Thus a bulk description of black hole microstates is an open problem. Some recent developments by Witten \cite{Witten_2022} have suggested that in the large-\(N\) limit beyond the Hawking Page transition, black hole fluctuations are described via type III algebra. Type III algebras have non-trivial commutators, suggesting that this transition can be interpreted as the emergence of the black hole interior. There is strong evidence that the large-\(N\) limit could be the answer to understanding the black hole interior at the finite-\(N\) limit. The description of the Yang-Mills at finite-\(N\) is exceptionally non-trivial. However, it may be fruitful trying to understand the finite-\(N\)  of it's dual i.e. the bulk gravitational theory with finite-\(N\) corrections \cite{Burman:2023kko,2024arXiv240905850B,2024JHEP...03..162K}\\ \\ The upshot of using the machinery of Virasoro blocks is that it makes possible to construct the full spectrum of the quasi-normal modes of the black hole, including the \(1/N^2\). Motivated by finite-\(N\) effects, our goal is to understand the heavy conformal blocks at finite \(c\), and how they lead to information restoring effects.
\chapter*{Outline}
\begin{large}
\textbf{Chapter 2} 
\end{large}
\vspace{2mm}
\\ In recent times, we have seen a remarkable surge in the conformal bootstrap approach to solving Conformal Field Theories (CFT)s \cite{Polyakov:1974gs,BELAVIN1984333}, with great success in the phenomenological and theoretical import \cite{Rychkov_2009,Heemskerk_2009,Heemskerk_2010,Poland_2011}. Alternatively, there has also been massive progress in building the effective field theory (EFT) in AdS and its studying it's analogous CFT dual. These scientific breakthroughs has lead to a bottom's up classification of all CFTs that have dual descriptions as EFTs in AdS. Furthermore, this technique holds merit in studying bootstrap in Mellin space using the perturbative scattering amplitudes \cite{Fitzpatrick_2012}. \\ \\ 
\begin{large}
\textbf{Chapter 3 - 4} 
\end{large}
\vspace{2mm}
\\ Lately conformal bootstrap has led to a rigorous, non-perturbative proof of the cluster decomposition principle in AdS\(_{d+1}\) for all unitary CFTs in \(d \geq 3\) \cite{Fitzpatrick_2013}. The AdS cluster decomposition with all the perturbative corrections is deeply linked to the operator content of an OPE in its dual CFT. We know from the first principles of conformal field theories that states in CFT\(_2\) are allowed to be written in terms into irreducible representations of the Virasoro blocks. As a consequence, all CFT correlators are represented as sum over the exchange of Virasoro conformal blocks \cite{Heemskerk_2009,Heemskerk_2010,Fitzpatrick_2013, Fitzpatrick_2014}. Via AdS/CFT, one can view these Virasoro blocks as an exchange over the sum of AdS wavefunctions, with physical states in the Virasoro block corresponding to the wavefunctions of some primary operators in AdS\(_3\) along with the intermediate exchange gravitons. The punchline being that these Virasoro blocks capture the quantum effects of gravity in AdS\(_3\). Another interesting application of Virasoro blocks has been in studying quantum entanglement \cite{Ryu_2006,Ryu_2006*} and scrambling \cite{Roberts_2015,Fitzpatrick:2016thx}, both in the vacuum and in the background of a heavy pure state of AdS. Recent work has been done by using modular invariance to study the spectrum of CFT\(_2\) at large central charge \cite{Hartman_2014}.
\\ \\
\begin{large}
\textbf{Chapter 5} 
\end{large}
\vspace{2mm}
\\ There is an elegant prescription for representing four-point conformal blocks with external scalar operators as \enquote{geodesic Witten diagrams} \cite{Hijano_2015, Hijano_2016,Penedones_2011}. The geodesic variant of a Witten diagram is the usual Witten diagram but with vertices integrated over geodesics connecting the external operators, rather than over all of AdS spacetime. \\ \\
\begin{large}
\textbf{Chapter 6 - 7} 
\end{large}
\vspace{2mm}
\\ Black hole thermodynamics naturally comes out of the symmetry algebra of AdS\(_3\)/CFT\(_2\) \cite{Fitzpatrick_2014,Fitzpatrick_2015,Fitzpatrick_2016,Alkalaev_2015,Hijano_2015,Beccaria_2016,Besken_2016,fitzpatrick2016conformalblockssemiclassicallimit}. Information loss occurs in the semi-classical limit (large \(N\) or large \(c\)), however this can be resolved by incorporating non-perturbative \(e^{-c}\) effects \cite{fitzpatrick2016informationlossads3cft2, Chen_2017*}. These observations enforces the possibility that in \(2+1\) dimensions, black hole thermodynamics, information loss, and restoration are completely dependent only on the Virasoro blocks. This suggests that quantum gravity in AdS\(_3\) maybe independent of the spectrum and OPE coefficients of the dual CFT\(_2\). This feature helps us exploit the tools of non-perturbative effects in Virasoro blocks to understand the physics in quantum gravity.
\chapter{Review of Conformal Bootstrap}
\begin{figure}[h]
    \centering
    \includegraphics[width=0.30\linewidth]{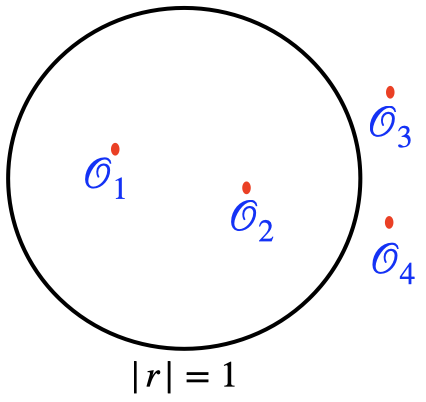}
    \caption{Radial quantisation in CFTs}
    \label{fig:radqfig}
\end{figure}

A conformal field theory (CFT) is characterised by a infinite set of local operators \(\mathcal{O}_{1}(x),\mathcal{O}_{2}(x),...\) \cite{Poland2016,Poland_2019} which capture the data of the physical system at the point \(x\). Suppose we have an equal-time n-point correlation function of local primaries,
\begin{equation}
    \braket{\mathcal{O}_{1}(x_1)...\mathcal{O}_{n}(x_n)}
\end{equation}
where the operator insertions, \(x_i\) are embedded in \(\mathbb{R}^d\). For example, in the Ising model, we measure the spin \(s_i\) at the lattice point \(i\). In the continuum limit, we can view the lattice as a continuous space \(i \rightarrow x\), with \(x \in \mathbb{R}^d\). The correlation function of \(s_i\) with other spins becomes a smooth function of \(x\), defining the spin operator \(\sigma(x)\). Alternatively we could have started with the local energy density \(\sum_{|i-j|=1}s_i s_j\), which would have given us the energy operator \(\epsilon(x)\). \\ Conformal symmetry severely constraints the correlation function of the local primary operators. For example, the two- and three-point functions of scalar primaries are given by,
\begin{align}
    &\braket{\mathcal{O}(x_1)\mathcal{O}(x_2)}=\frac{1}{|x_1 -x_2|^{2\Delta_{\mathcal{O}}}} \nonumber \\ &\braket{\mathcal{O}(x_1)\mathcal{O}(x_2)\mathcal{O}(x_3)}=\frac{f_{123}}{|x_1 -x_2|^{\Delta_1 +\Delta_2-\Delta_3}|x_3-x_1|^{\Delta_1+\Delta_3-\Delta_2}|x_2-x_3|^{\Delta_2+\Delta_3-\Delta_1}}
\end{align}
where \(\Delta_{\mathcal{O}}\) is the scaling dimension of the primary operator \(\mathcal{O}\). In a general setting, these scalar operators can transform non-trivially under rotations (carry spin \(l_{\mathcal{O}}\)), which modifies the numerator \cite{Osborn_1994}. Scaling dimensions are closely related to the critical exponents which can be measured in experiments. For instance, in the case of the Ising model, we have critical exponents \(\nu\) and \(\gamma\) which describe the behaviour of the correlation length \(\xi\) and susceptibility \(\chi\) as one approaches the critical temperature \(T_c\),
\begin{align}
    &\xi \propto |T-T_c|^{-\nu} \nonumber \\ 
    &\chi \propto |T-T_c|^{-\gamma}
\end{align}
In d=2, the critical exponents of the Ising model can be calculated from minimal models. These exponents are given by,
\begin{align}
    &\nu=\frac{1}{d-\Delta_{\epsilon}}=1 \nonumber \\ &\gamma=\frac{2-2\Delta_{\sigma}}{2-\Delta_{\epsilon}}=\frac{7}{4}
\end{align}
Conveniently, the scaling dimensions \(\Delta_{\mathcal{O}}\), spins \(l_{\mathcal{O}}\) and the three-point OPE coefficients \(f_{123}\) is enough to determine any correlation function in the theory. In the 1970's, Ferrara, Gatto, Grillo, and
Polyakov discovered a way of gluing three-point functions into higher point functions, usually known as the conformal block decomposition. For example, the four-point function of \(\sigma\)s can be be obtained by gluing three-point functions \(\braket{\sigma \sigma \mathcal{O}}\) and \(\braket{\mathcal{O} \sigma \sigma}\) and summing over the operators \(\mathcal{O}\),
\begin{figure}[h]
    \centering
    \includegraphics[width=0.45\linewidth]{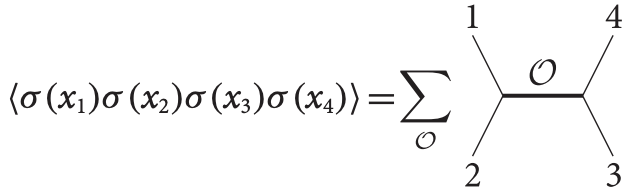}
\end{figure} \squeezeup
\begin{equation}
    =\sum_{\mathcal{O}}f^2_{\sigma \sigma \mathcal{O}} g_{\Delta_{\mathcal{O}},l_{\mathcal{O}}}(x_1,x_2,x_3,x_4)
\end{equation}
The factor \(f^2_{\sigma \sigma \mathcal{O}}\) is the product of the two OPE coefficients that we get from gluing the two three-point functions, whereas \(g_{\Delta_{\mathcal{O}},l_{\mathcal{O}}}\) is the conformal partial wave or commonly referred as the conformal block. \\ Hence, the conformal symmetry reduces the CFT variables to a set of numbers \((\Delta_{\mathcal{O}},l_{\mathcal{O}},f_{123})\). We can compute the same four-point correlator, however this time with a different choice of gluing,
\begin{figure}[H]
    \centering
    \includegraphics[width=0.45\linewidth]{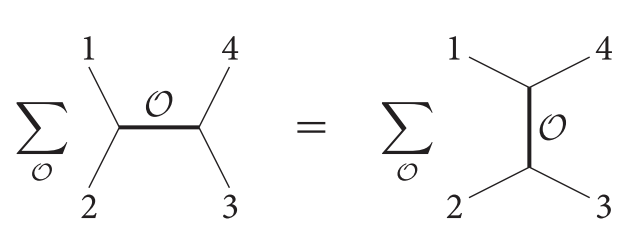}
    \caption{This figure illustrates the crossing-symmetry of the four-point correlator.}
    \label{fig: crossing}
\end{figure}
The central theme of the conformal bootstrap is that by imposing the crossing relation, we should be able to significantly winnow down the set of all possible CFT data. Ideally, if we impose crossing relations for all 4pt functions of the theory, we will be left with the CFT data corresponding to the actually existing critical theories. Unfortunately in practice, it has so far been possible to impose crossing relations on only a handful of 4pt functions at a time. Nonetheless, in the subsequent chapters, we will see that even this limited procedure imposes very non-trivial constraints in CFTs. \\ In a conformal field theory, any four-point correlation function can be expanded in terms of a conformal partial wave \cite{dolan2012conformalpartialwavesmathematical,Dolan_2001,Dolan_2004, Simmons_Duffin_2014} which are commonly known as conformal blocks \cite{Fitzpatrick_2013,Simmons_Duffin_2017}. This expansion is equivalent to the S-matrix calculations we often do in field theory. In the radial quantisation, we are able to define any four-point function as a product of two two-point functions,
\begin{equation}
\braket{\mathcal{O}_1(x_1)\mathcal{O}_2(x_2)| \mathcal{O}_3(x_3)\mathcal{O}_4(x_4)}
\end{equation}
living on a sphere separating \(x_1,x_2\) and \(x_3,x_4\) as illustrated by Figure \ref{fig:radqfig}
The CPW then corresponds to inserting of a projection operator \(\mathcal{P}_{\Delta,l}\), where the projector will be given by,
\begin{equation}
    \mathcal{P}_{\Delta,l}=\sum_{\alpha, \beta= \mathcal{O}, P_{\mu}\mathcal{O}, P_{\mu}P_{\nu}\mathcal{O},...} \ket{\alpha} \mathcal{N}_{\alpha \beta}^{-1}\bra{\beta}
\end{equation}
where \(\mathcal{N}_{\alpha \beta}=\braket{\alpha | \beta}\) is the \textit{Gram matrix} of the multiplet. Then the four-point OPE becomes,
\begin{equation}
    \braket{\mathcal{O}_1(x_1)\mathcal{O}_2(x_2)[\mathcal{P}_{\Delta,l}]\mathcal{O}_3(x_3)\mathcal{O}_4(x_4)}=\frac{g(u,v)}{x_{12}^{2\Delta_{\mathcal{O}}}x_{34}^{2\Delta_{\mathcal{O}}}}
\end{equation}
and
\begin{equation}
    g(u,v)=\sum_{\Delta,l} P^2_{\Delta,l}\ g_{\Delta,l}(u,v)
\end{equation}
where \(g_{\Delta,l}(u,v)\) is the conformal block written in terms of the cross-ratios \(u\) and \(v\),
\begin{equation}
    u = \left(\frac{x_{12}x_{34}}{x_{24}x_{13}}\right)^2  \;\ \text{and} \;\ v=\left(\frac{x_{14}x_{23}}{x_{24}x_{13}}\right)^2 
\end{equation}
The function \(P_{\Delta,l}\) corresponds to the contribution of the conformal descendants associated to the primary operator \(\mathcal{O}\), and it's completely fixed by demanding conformal invariance. However, the coefficients of the OPE are theory-dependent and in general cannot be determined. \\ The expansion of the primary operators can be done in any of the three channels defined using Mandelstam variables (s,t,u) and it's this equivalence that is usually refereed to as crossing symmetry equation (displayed in Figure \ref{fig: crossing}). The two most common examples found in literature is the \(\mathcal{O}_1\mathcal{O}_2\) and \(\mathcal{O}_3\mathcal{O}_4\) contractions known as the s-channel, while the \(\mathcal{O}_1\mathcal{O}_3\) and \(\mathcal{O}_2\mathcal{O}_4\) contractions is called the t-channel. The crossing relation equates these two correlation functions,
\begin{equation}
    \sum_{\Delta,l} P_{\Delta,l}\ g_{\Delta,l}(u,v)= \left(\frac{u}{v}\right)^{\Delta} \sum_{\Delta,l} P_{\Delta,l}\ g_{\Delta,l}(v,u) \label{cb}
\end{equation}
There is an alternative representation of CPWs due to Dolan and Osborn \cite{Dolan_2004}. We know that the conformal algebra group is \(SO(d+1,1)\) with generators \(L_{ab}\). The Casimir \(C=1\frac{1}{2}L_{ab}L^{ab}\) acts with the same eigenvalue on every state in an irreducible representation. 
\begin{align}
    &C\ket{\mathcal{O}}= \lambda_{\Delta, l}\ket{\mathcal{O}} \nonumber \\ &\lambda_{\Delta, l}=\Delta(\Delta-d)+l(l+d-2)
\end{align}
Let \(\mathcal{L}_{ab,i}\) be a differential operator giving the action of the conformal generator \(L_{ab}\) on the operator \(\mathcal{O}(x_i)\)
\begin{align}
    (\mathcal{L}_{ab,1}+\mathcal{L}_{ab,2})\mathcal{O}(x_1)\mathcal{O}(x_2)\ket{0}= &\left([L_{ab}\mathcal{O}(x_1)]\mathcal{O}(x_2)+\mathcal{O}(x_1)[L_{ab},\mathcal{O}(x_2)]\right)\ket{0} \nonumber \\ = &L_{ab}\mathcal{O}(x_1)\mathcal{O}(x_2)\ket{0}
\end{align}
Thus, the Casimir now can be written as a differential operator
\begin{equation}
    C\mathcal{O}(x_1)\mathcal{O}(x_2)\ket{0}=\mathcal{D}_{1,2}\mathcal{O}(x_1)\mathcal{O}(x_2)\ket{0}
\end{equation}
We find that the conformal block \(g_{\Delta,l}\) satisfies the differential equation,
\begin{equation}
    \mathcal{D}g_{\Delta,l}=\lambda_{\Delta, l}g_{\Delta,l}
\end{equation}
where the second order differential operator is,
\begin{align}
    \mathcal{D} \equiv &\sum_{ab}(\mathcal{L}_{ab}^{(1)}+\mathcal{L}_{ab,2}^{(2)}) \nonumber \\ = &[z^2 (1-z)\partial_{z}^{2}-z^2 \partial_z +(z \rightarrow \bar{z})]+(d-2)\frac{z\bar{z}}{z-\bar{z}}[(1-z)\partial_z-(z \rightarrow \bar{z})]
\end{align}
By solving the Casimir equation we are able to find the structure of the conformal block. For example in 2d and 4d we have exact results,\cite{Dolan_2001,Dolan_2004}
\begin{align}
    &g_{\Delta,l}^{(2d)}=k_{\Delta+l}(\bar{z})k_{\Delta-l}(z)+(z \rightarrow \bar{z}) \\  &g_{\Delta,l}^{(4d)}=\frac{z\bar{z}}{z-\bar{z}}\left(k_{\Delta+l}(\bar{z})k_{\Delta-l-2}(z)-(z \rightarrow \bar{z})\right) \\ &k_{\beta}(x)=x^{\frac{\beta}{2}}\ _2F_1 \left(\frac{\beta}{2},\frac{\beta}{2},\beta;x \right)
\end{align}
In odd dimensions, no explicit formulation of these functions is known. However, the conformal blocks can still be computed in odd dimensions using a technique known as recursion relations. 
\section{Bootstrap for Mean Field Theories}
Now let's consider a four-point correlators for a mean field theory (MFT), where all correlators are represented as a product of two-point wick contracted correlators. An advantage of working with MFTs is that they can be defined from free quantum field theories dual to AdS. Let's consider our usual four-point correlator of identical fields, with the contraction (\(12 \rightarrow 34\)):
\begin{equation}
    \braket{\mathcal{O}(x_1)\mathcal{O}(x_2)\mathcal{O}(x_3)\mathcal{O}(x_4)}= \frac{1}{(x_{12}x_{34})^{2\Delta}} \sum_{\tau,l} P_{\tau,l}g_{\tau,l}(u,v)=\frac{1}{(x_{12}x_{34})^{2\Delta}}
\end{equation}
where we have expressed the four-point function in terms of the spin \(l\) which is even due to Bose symmetry, and a new quantity which we define as twist \(\tau=\Delta-l\). This tells us that in the s-channel, only the identity operator contributes in the OPE. However, the t-channel OPE happens to be non-trivial. Using equation \eqref{cb}, we see
\begin{equation}
    u^{-\Delta}=v^{-\Delta} \sum_{\tau,l} P_{\tau,l}g_{\tau,l}(v,u) \label{mft}
\end{equation} 
Working in the limit \(u \ll v \ll 1\), often called the \enquote{lightcone limit}. However, in this limit there appears to be a simple problem in \eqref{mft}, i.e. there is a power-law singularity \( u^{-\Delta}\) on the LHS, whereas there is a \(\log u\) on the RHS, and naively these two don't seem to match. A simple resolution comes to the rescue to this paradox; the sum over conformal blocks does not converge uniformly near \(u=0\). The sum converges for real and positive \(u\) when, but is divergent \(\mathrm{Re}(\sqrt{u})<0\). So we must define the sum over conformal
blocks for general \(u\) as the analytic continuation of the sum in the convergent region. The key to matching the two sides is noticing that 
\begin{equation}
    [\mathcal{O}\mathcal{O}]_{n,l} \sim \mathcal{O} \partial^{\mu_1}\partial^{\mu_2}....\partial^{2n}\mathcal{O}
\end{equation}
 \(\tau_1+\tau_2+2n\). These operators are called \enquote{\textit{double-twist} operators}, and represent an infinite family of conformal operators with increasing spin \(l\) and twist approaching \(\tau_1+\tau_2+2n\). The resolution to this problem, as discussed in \cite{Fitzpatrick_2013} is, for the bootstrap equation to be satisfied, we require that the sum over the conformal blocks in \label{mft} not to converge uniformly in \(u\) and \(v\). In order to understand this series, we need to study the conformal blocks at large \(l\) or \(\tau\), in the lightcone limit (\(u \ll v \ll 1\)), in the regime where \(l \rightarrow \infty\) with \(l\sqrt{u}\) fixed. In the lightcone limit, with large \(\tau\), the conformal blocks are always suppressed by an overall twist factor \(u^{\frac{\tau}{2}}\) or \(v^{\frac{\tau}{2}}\), and the conformal block coefficients are bound at large \(\tau\) \cite{Pappadopulo_2012}. This implies that at small \(|u|\) and \(|v|\), the sum over \(\tau\) converges. \\ We concentrate our efforts to the large \(l\) limit with fixed \(\tau\) for the light-cone limit OPE. The t-channel conformal block for a four-point correlation function can be written as 
 \begin{equation}
     \braket{\mathcal{O}(x_1)\mathcal{O}(x_2)\mathcal{O}(x_3)\mathcal{O}(x_4)}=\frac{1}{(x_{14}x_{23})^{2\Delta}} \sum_{\tau,l} P_{\tau,l}g_{\tau,l}(v,u)
 \end{equation}
 \begin{figure}
     \centering
     \includegraphics[width=0.80\linewidth]{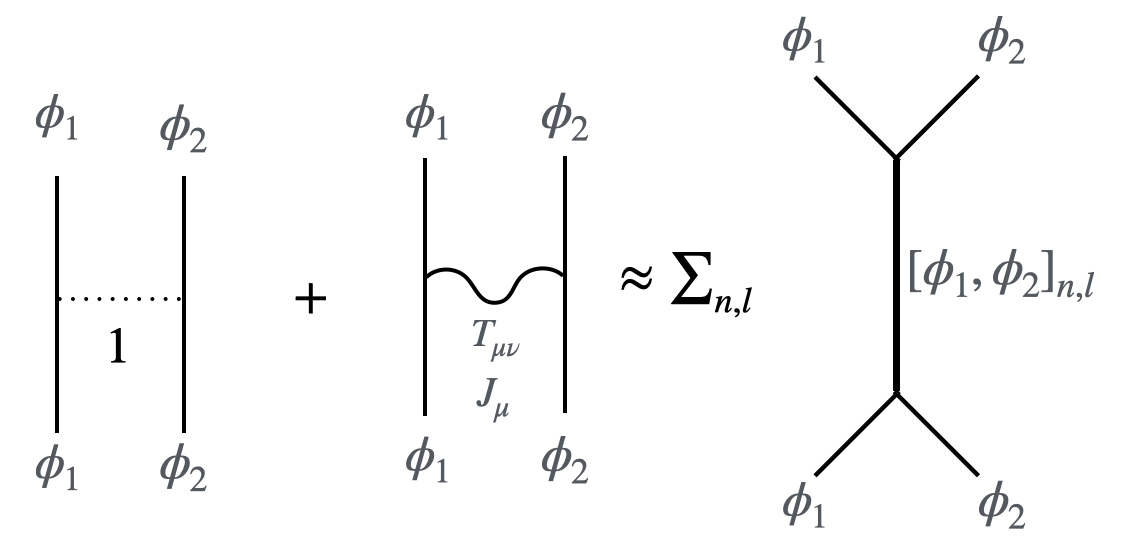}
     \caption{This figure suggests the form of the Conformal Bootstrap equation in the lightcone OPE limit where the conformal cross-ratio \(u \rightarrow 0\). The first term of the left-hand side i.e. the Virasoro vacuum block is the dominant term comes from the exchange of the identity operator. The contribution of right-hand side comes from the exchange of double-twist operators.}
     \label{fig:enter-label}
 \end{figure}
 Specifically for \(d=2\), the global conformal blocks in this expansion take on a convenient form,
 \begin{equation}
    g_{\tau,l}(v,u)=k_{\tau+2l}(1-z)k_{\tau}(1-\bar{z})+k_{\tau+2l}(1-\bar{z})k_{\tau}(1-z) 
 \end{equation}  
 where \(u=z\bar{z}\), \(v=(1-z)(1-\bar{z})\) and \(k_{2\beta}(x)=x^{\beta} _2F_1(\beta,\beta,2\beta;x)\). We consider the regime \((1-\bar{z})< 1\), where the second term can be ignored as its exponentially suppressed in large \(l\). The hypergeometric function \(k_{\tau+2l}(1-z)\) has an integral representation as follows:
 \begin{equation}
     k_{\tau+2l}(1-z)=\frac{\Gamma(2l+\tau)}{\Gamma(l+\frac{\tau}{2})^2} \int_0^1 \frac{dt}{t(1-t)}\left(\frac{(1-z)t(1-t)}{1-t(1-z)}\right)^{\frac{\tau}{2}+l} 
 \end{equation}
 At large \(l\), the \(l\)-dependent piece in the integral has a sharp peak at \(t_*=\frac{1-\sqrt{z}}{1-z}\), while the \(\tau\)-dependent piece varies slowly around this peak and behaves like 
 \begin{equation}
     \left(\frac{(1-z)t(1-t)}{1-t(1-z)}\right)^{\tau} \sim 1+O(\sqrt{z})
 \end{equation}
 Now we will use the Stirling approximation to manipulate the \(\Gamma\) functions 
 \begin{align}
     &\Gamma(x)=\sqrt{\frac{2\pi}{x}}\left(\frac{x}{e}\right)^x\left(1+O\left(\frac{1}{x}\right)\right) 
 \end{align}
 which gives 
 \begin{equation}
     k_{2l+\tau}=2^{\tau}k_{2l}\left(1+O(\sqrt{z})\right)
 \end{equation}
 In the small \(z\) limit, which is equivalent to \(u\) we find 
 \begin{equation}
     g_{\tau,l}(v,u)=2^{\tau}k_{2l}(1-z)k_{\tau}(v)
 \end{equation}
 The function \(k_{2l}(1-z)=(1-z)^l \ _2F_1(l,l,2l;1-z)\) can further be simplified, if we consider working in the \(l \rightarrow \infty\) with \(y \equiv zl^2\) fixed such that \(y \lesssim O(1)\). In that case, we get
 \begin{equation}
     \frac{\Gamma(l)^2}{\Gamma(2l)}\ _2F_1(l,l,2l;1-z)=\int_0^1 \frac{t^{l-1}}{(1-t)}e^{\frac{-ty}{(1-t)l}}\left(1+O\left(\frac{1}{l}\right)\right)
 \end{equation}
 The integral is greatly simplified, after a simple change of variables \(s\equiv \frac{t}{1-t}\)
 \begin{align}
      \frac{\Gamma(l)^2}{\Gamma(2l)}\ _2F_1(l,l,2l;1-z) = &\int_0^{\infty}\frac{ds}{s}e^{-\frac{sy}{l}-\frac{l}{s}}\left(1+\mathcal{O}\left(\frac{1}{l}\right)\right) \nonumber \\ \approx &2K_0(2\sqrt{y})
 \end{align}
 where \(K_0\) is the modified Bessel function of the second kind. \\ Hence, we see that in this limit, the global conformal block at fixed \(\tau\) and large \(l\) takes the form 
 \begin{equation}
     g_{\tau,l}(v,u) \approx 2^{\tau+2l}v^{\frac{\tau}{2}}\sqrt{\frac{l}{\pi}}K_0(2\sqrt{u})
 \end{equation}
 We see that at small \(v\),the lowest twist terms \((n=0)\) dominate. In a MFT, the conformal block coefficient in any dimension \(d\) \cite{Fitzpatrick_2012} is,
 \begin{equation}
     P_{\tau+2n,l}= \frac{(1+(-1)^l)(\Delta-\frac{d}{2}+1)_n^2(\Delta)^2_{n+l}}{l! n!(l+\frac{d}{2})_n(2\Delta+n-d+1)_n(2\Delta+2n+l-1)_l(2\Delta+n+l-\frac{d}{2})_n}
 \end{equation}
 where \((a)_b\) is the Pochhammer symbol. For \((n=0)\) and large even \(l\) will be, we can approximate
 \begin{equation}
     P_{2\Delta,l} \stackrel{l \gg 1}{\approx} \frac{64\sqrt{\pi}}{\Gamma(\Delta)^2}\frac{l^{2\Delta-\frac{3}{2}}}{2^{2\Delta+2l}}
 \end{equation}
 Thus \eqref{mft} now becomes 
 \begin{equation}
     v^{-\Delta} \sum_{n,\text{large}\ l } P_{\tau_n,l}g_{\tau,l}(v,u) \approx \frac{1}{\Gamma(\Delta)^2} \sum_{\text{large even } \ l }^{\infty}l^{2\Delta-1}K_0(2\sqrt{u})
 \end{equation}
 The sum can be converted into an integral at large \(l\), and the result exactly produces the \(u^{-\Delta}\) power-law singularity on the LHS of \eqref{mft}. Furthermore, it replicates the \(u \rightarrow 0\) behaviour, just as we desire.
 \section{Heavy-Light Virasoro blocks}
 It is well establish that in a CFT, any correlation functions can be represented as a sum over conformal blocks. These conformal blocks play a key role in establishing the CFT bootstrap relation, and as a consequence, obtaining analytic constraints on CFT. Using the AdS/CFT correspondence, we can re-interpret the Virasoro conformal blocks as the exchange over gravitons or other two-particle bound states in AdS\(_3\). Essentially, the Virasoro blocks capture quantum gravitational effects in AdS\(_3\).
 \section{HHLL correlators in AdS}
 We will understand Virasoro conformal blocks in the limit of large central charge \(c=\frac{3}{2G} \ll 1\), commonly also known as the semi-classical limit. This is analogous to working with large-\(N\) limit of gauge theories. In the CFT, the bulk field is dual to light primary operator \(\mathcal{O}_L\) with conformal dimension \(\Delta_L\), where \(\frac{\Delta_L}{c} \rightarrow 0\) as \(c \rightarrow \infty\). Meanwhile, the heavy CFT operators \(\mathcal{O}_H\) create conical defects when their conformal dimension \(\Delta_H \sim O(c)\) is less than the critical BTZ threshold \( \Delta_*=\frac{c}{24}\). \\ Using this perspective, heavy AdS backgrounds, the four-point correlation takes the form 
 \begin{equation}
    \braket{\mathcal{O}_H(0)\mathcal{O}_L(\infty)\mathcal{O}_L(1)\mathcal{O}_H(z)}
 \end{equation}
 Our next task will be to represent the Virasoro conformal block for a four-point heavy-light correlator in terms of a global conformal block using a new set of coordinates.
 \section{Light Virasoro blocks at large \(c\)}
 Before moving onto the working of heavy-light Virasoro blocks, let us review how light Virasoro blocks behave in the semi-classical limit with other operator dimensions fixed
 \begin{equation}
    \braket{\mathcal{O}_L(0)\mathcal{O}_L(\infty)\mathcal{O}_L(1)\mathcal{O}_L(z)}
 \end{equation}
 with \(c \rightarrow \infty\) and dimension \(h_L\) fixed. \\ The global block takes a simple form 
 \begin{equation}
     G_h(z)=(1-z)^{h-2h_L}\ _2F_1(h,h,2h;1-z)+O\left(\frac{1}{c}\right)
 \end{equation}
 where the intermediate exchange operators have conformal weight \(h\), which too also fixed in the \(c \rightarrow \infty\) limit. Thus the Virasoro conformal block reduces to a global block, generated by the global Virasoro algebra generators \(L_0\) and \(L_{\pm}\) in the large \(c\) limit.\\ Let's consider a projection operator \(\mathcal{P}_h\) onto the states of the Virasoro block, with the basis
 \begin{equation}
     L^{k_1}_{-m_1}....L^{k_n}_{-m_n}\ket{h}
 \end{equation}
 and the complete projection operator is defined as 
 \begin{equation}
     \mathcal{P}_h \approx \sum_{\{m_i,k_i\}}\frac{L^{k_1}_{-m_1}....L^{k_n}_{-m_n}\ket{h} \bra{h}L^{k_1}_{m_1}....L^{k_n}_{m_n}}{\mathcal{N}_{ \{m_i,k_i \}}}
 \end{equation}
 \begin{figure}
     \centering
     \includegraphics[width=0.95\linewidth]{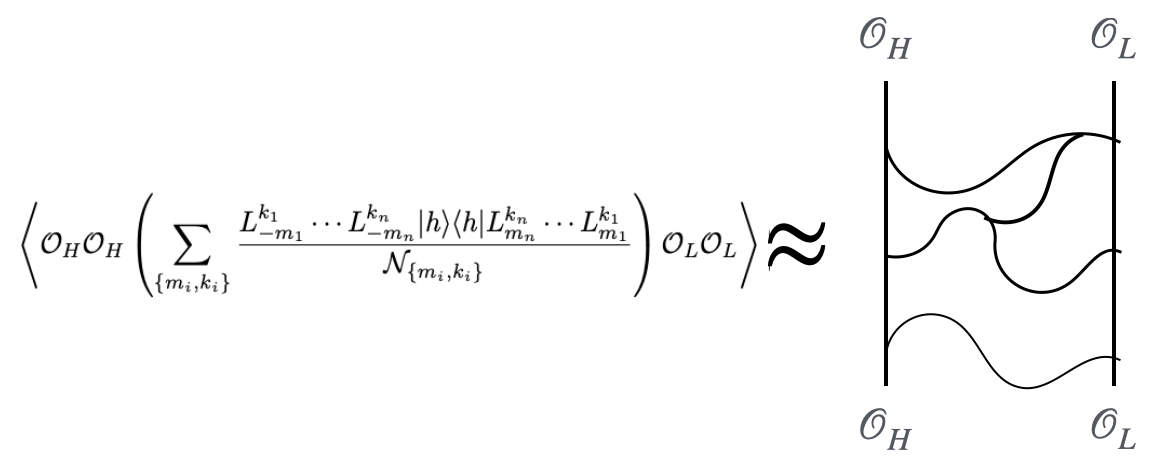}
     \caption{This figure suggests how the exchange of all descendants of the identity operator in the Virasoro algebra corresponds to the exchange of all multi-graviton states in AdS\(_3\). This will be sufficient in building the full non-perturbative AdS\(_3\) gravitational fields entirely from CFT\(_2\). }
     \label{fig:enter-label}
 \end{figure}
 The projection operator sums over all the intermediate states created by the action of the Virasoro generators on the states \(\ket{h}\). The normalisation is determined using 
 \begin{equation}
     \bra{h}L_nL_{-n}\ket{h}=\left(2nh+\frac{c}{12}n(n^2-1)\right)
 \end{equation}
 for \(n>0\)\, and so for\(n \geq 2\), the normalisation is of \(O(c)\). The action of \(L_{-m}\) in the numerator of \(\mathcal{P}_h\) can produce powers \(h\) and \(h_L\) using
 \begin{equation}
     [L_m, \mathcal{O}(z)]=h_i(1+m)z^m\mathcal{O}(z)+z^{1+m}\partial_z \mathcal{O}(z)
 \end{equation}
 The numerator of \(\mathcal{P}_h\) is independent of any powers of \(c\). Hence, we can write the projection operator as 
 \begin{equation}
     \mathcal{P}_h \approx \sum_k \frac{L_{-1}^k\ket{h}\bra{h}L_1^k}{\bra{h}L_{1}^kL_{-1}^k\ket{h}} + O\left(\frac{1}{c}\right)
 \end{equation}
 This just produces the global block for the holomorphic sector.
 \section{Heavy-Light Virasoro blocks at large \(c\)} \label{Heavy-light blocks}
 Now we will get back to computing the four-point HHLL correlator 
 \begin{equation}
     \mathcal{V}(z)=\braket{\mathcal{O}_H(\infty)\mathcal{O}_H(1)[\mathcal{P}_h]\mathcal{O}_L(z)\mathcal{O}_L(0)}
 \end{equation}
 \begin{figure}[h]
     \centering
     \includegraphics[width=0.35\linewidth]{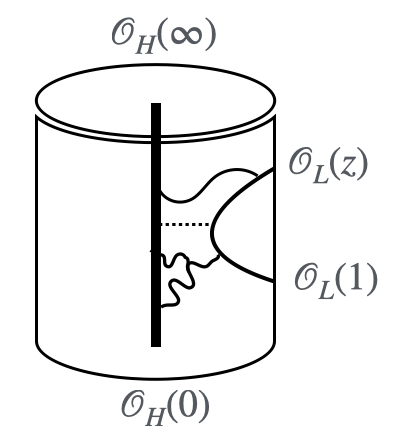}
     \caption{This figure depicts a heavy-light CFT correlator and a light operator interacting with a deficit angle or BTZ black hole.}
     \label{fig:enter-label}
 \end{figure}
 \\ and \(\mathcal{P}_h\) projects onto \(\ket{h}\) and all it's descendants. To compute this Virasoro block, we will change coordinates from \(z\) to \(w\). The Virasoro block written in terms of \(w\) 
 \begin{equation}
     \mathcal{V}(w)=\braket{\mathcal{O}_H(\infty)\mathcal{O}_H(1)[\mathcal{P}_h]\mathcal{O}_L(w)\mathcal{O}_L(0)}
 \end{equation}
 differs from \(\mathcal{V}(z)\) by a Jacobian factor of \((z'(w))^{h_1}(z'(0))^{h_2}\). The projection operator is now defined as 
 \begin{equation}
     \mathcal{P}_{h,w} \approx \sum_{{m_i,k_i}}\frac{\mathcal{L}^{k_1}_{-m_1}....\mathcal{L}^{k_n}_{-m_n}\ket{h} \bra{h}\mathcal{L}^{k_1}_{m_1}....\mathcal{L}^{k_n}_{m_n}}{\mathcal{N}_{m_i,k_i}}
 \end{equation}
 where the new Virasoro generators\(\mathcal{L}\)'s are related to the old \(L\)'s using the conformal transformation property of the stress tensor 
 \begin{equation}
     \sum_n w^{-n-2}\mathcal{L}_n= \left(\frac{dz}{dw}\right)^2\sum_m z^{-m-2}L_m+\frac{c}{12}S(z(w),w)
 \end{equation}
 The only non-trivial component of this calculation is computing the heavy-matrix elements
 \begin{equation}
     \braket{\mathcal{O}_H(\infty)\mathcal{O}(1)\mathcal{L}^{k_1}_{-m_1}....\mathcal{L}^{k_n}_{-m_n}|h}
 \end{equation}
 Let's make a choice for the new coordinate \(w(z)\)
 \begin{align}
     &1-w=(1-z)^{\alpha} \;\ \text{where} \;\ \alpha=\sqrt{1-24\frac{h_H}{c}} \label{wz}
 \end{align}
 \\ Note that \(\alpha\) is related to the Hawking temperature \(T_H\) of a BTZ black hole 
 \begin{equation}
     T_H=\frac{|\alpha|}{2\pi} \;\ \text{when} \;\ h_H > \frac{c}{24}
 \end{equation}
 In the case of global generator \(\mathcal{L}_{-1}\), the computation simplifies a lot
 \begin{align}
     \braket{\mathcal{O}_H(\infty)\mathcal{O}(1)\mathcal{L}^k_{-1}|h}=&\lim_{w \rightarrow 0}\partial_w^k\braket{\mathcal{O}_H(\infty)\mathcal{O}(1)\mathcal{O}_w} \nonumber \\ &=\alpha^{-h}C_{HHh}\lim_{w \rightarrow 0}\partial_w^k(1-w)^{-h}
 \end{align}
 and \(C_{HHh}\) is the OPE coefficient. The leading order behaviour of the OPE will be 
 \begin{equation}
    \braket{\mathcal{O}_H(\infty)\mathcal{O}_H(1)\left(\sum_k \frac{\mathcal{L}_{-1}^k\ket{h}\bra{h}\mathcal{L}_1^k}{\bra{h}\mathcal{L}_{1}^k\mathcal{L}_{-1}^k\ket{h}}\right)\mathcal{O}_L(w)\mathcal{O}_L(0)}
 \end{equation}
 Hence, the Virasoro block becomes
 \begin{equation}
     \mathcal{V}(c,h_p,h_i,z)=(1-w)^{h_L(1-\frac{1}{\alpha})}\left(\frac{w}{\alpha}\right)^{h-2h_L}\ _2F_1(h,h,2h;w) \label{vira}
 \end{equation}
 where \(w\) and \(z\) are related via \eqref{wz}, and we have included the contributions from the Jacobian factor \(w'(z)^{h_i}\).
 \chapter{Deficit angles in 3d Gravity}
 In (2+1) dimensional gravity, there are no local propagating degrees of freedom. This implies that every solution to pure 3d gravity with a negative cosmological constant is locally isometric to the empty anti-de Sitter (AdS) spacetime. At a semi-classical level, black holes are characterized by a Hawking temperature, and in AdS/CFT \cite{witten1998antisitterspaceholography,Gubser_1998,banks1998adsdynamicsconformalfield} are dual to thermal states of the boundary CFT theory.
 \begin{figure}
     \centering
     \includegraphics[width=0.70\linewidth]{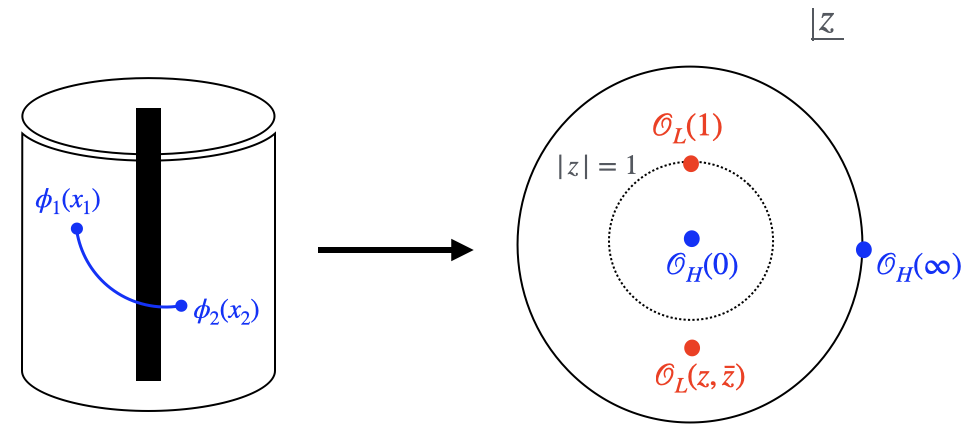}
     \caption{(Left) The figure depicts the HHLL correlator in AdS with a non-trivial background geometry. (Right) using radial quantisation a dual CFT is depicted where the background is created by the two heavy operators.}
     \label{fig:enter-label}
 \end{figure}
 The only physical class of solutions are AdS\(_3\) and its quotients, BTZ black hole being the most obvious one, however other solutions have been obtained as well \cite{Edery_2021,Edery:2022crs}. BTZ black holes need to lie above a critical mass threshold of \(\Delta_*=\frac{c}{24}=\frac{1}{8G}\) to exist, separating them from the pure AdS\(_3\) vacuum. An important observation is that below \(\Delta_*\), no black hole like geometry exists in pure gravity. Einstein's solutions are known to have non-trivial solutions in the presence of sources \cite{DESER1984220,DESER1984405}. In particular, a point-like source of sub-planckian mass placed in AdS\(_3\) will produce a conical defect at its location, while everywhere else the spacetime remains AdS\(_3\) locally. Additionally, in some recent work by Edery \cite{Edery_2021,Edery:2022crs}, he studied the solutions to Einstein's equations where a deficit angle exists without a conical singularity. \\
 In string theory and gravitational path integrals \cite{Lewkowycz_2013,benjamin2020puregravityconicaldefects}, the AdS\(_3/\mathrm{Z}_N\) play a crucial role. However, the parameter \(N\) which determines the defect’s strength can in general take non-integer values. It's important to note that most conical defect solutions are not quotients of AdS\(_3\). The metric of AdS\(_3\) with conical defect and BTZ black hole is the same, however, the horizon radius is analytically continued so that the mass falls below the BTZ mass threshold. This suggests that we can probe the black holes from below the threshold and connect heavy and thermal states.
 \section{Spectrum of states}
At the semi-classical level, the spectrum of primaries of a holographic 2d CFT dual to a pure AdS\(_3\) gravity comprises the identity and heavy primary operators \(\mathcal{O}_H\) whose conformal dimension lies above the BTZ threshold \(\Delta_*=\frac{c}{12}\). The Cardy density of these heavy operators which are dual to black holes is given by \cite{Hartman_2014}, and their conformal dimension is,
\begin{equation}
    \Delta_H=m=\frac{c}{12}(1-\alpha^2), \;\ \alpha=\sqrt{1-\frac{12\Delta_H}{c}}
\end{equation} \label{mass}
corresponds to the ADM mass of the BTZ black hole. To this universal part of the spectrum, we can add matter in two possible ways. The first way is to add conical defects to the AdS geometry so that the mass \(\Delta_H\) falls below the BTZ threshold and creates a particle \(\mathcal{O}_H\) of mass \eqref{mass}. The second way is to add a light scalar \(\mathcal{O}_L\) of conformal dimension \(\Delta_L \ll c\). The light operator is taken to be a generalized free field, so that it is dual in the bulk to the free scalar field \(\phi\) in the dual AdS. \\ We use both \(\mathcal{O}_L\) and \(\mathcal{O}_H\) to create double-twist primary operators. Suppose, we have two distinct primary operators \(\mathcal{O}_1\) and \(\mathcal{O}_2\), then we can construct an infinite family of double trace or \textit{double-twist} primaries that exists in an OPE construction,
\begin{equation}
    [\mathcal{O}_1\mathcal{O}_2]_{nl} \sim :\mathcal{O}_1 \Box^{n}\overleftrightarrow{\partial}^l\mathcal{O}_2:-(\text{traces})
\end{equation}
The trace subtraction is necessary to ensure that \([\mathcal{O}_1\mathcal{O}_2]_{nl}\)
is a primary operator. If the primaries \(\mathcal{O}_1\) and \(\mathcal{O}_2\) are identical, then the spin of the composite state \(l\) is even. The \textit{double-twist} primaries can be interpreted as two particle states whose bulk description is an orbiting configuration of \(\mathcal{O}_1\) and \(\mathcal{O}_2\), labelled by angular momentum \(l\) and energy by \(n\). Depending on \(\mathcal{O}_1\) and \(\mathcal{O}_2\) operator content, the \textit{double-twist operators} can be classified into three categories. It has been summarised in Table \ref{tab:my_label}.
\begin{table}[h]
    \centering
    \begin{tabular}{|c|c|c|c|}
    \hline
         Primary & Dimension \(\Delta\) &  Spin \(l\) & AdS interpretations \\
        \(\mathcal{O}\) & \(h+\bar{h}\) & \(h-\bar{h}\) & (Bulk excitation) \\ \hline
        \(\mathbb{1}\) & 0 & 0 & pure AdS vacuum \\  
        \(\mathcal{O}_L\) & \(\Delta=\Delta_L\) & 0 & free scalar field \\
        \(\mathcal{O}_L\) & \(\Delta_H=\frac{c}{12}(1-\alpha^2)\) & 0 & heavy background \\
         \([\mathcal{O}_L\mathcal{O}_L]_{nl}\) & \(\Delta_{nl}=2\Delta+2n+|l|\) & \(l \in 2\mathrm{Z}\) & light 2-particle state \\
         \([\mathcal{O}_H\mathcal{O}_L]_{nl}\) & \(\Delta_H+\alpha(\Delta+2n)+|l|\) & \(l \in \mathrm{Z}\) & orbiting light mode \\ \hline
    \end{tabular}
    \caption{The complete spectrum \((h,\bar{h})\) of primaries of a holographic CFT dual to an AdS spacetime.}
    \label{tab:my_label}
\end{table}
\\
\textbf{LL double twist fields}: If both \(\mathcal{O}_1\) and \(\mathcal{O}_2\) are light operators, then at large spin \(|l|\), the spectrum of \([\mathcal{O}_L\mathcal{O}_L]_{nl}\) approaches the universal MFT spectrum of \textit{double-twist} operators \cite{Fitzpatrick_2014}. The most trivial example of such a composite state is \([\mathcal{O}_L\mathcal{O}_L]_{00}=:\mathcal{O}^2_L:\)\\
\textbf{HL double twist fields}: The HL-double twist operators are \([\mathcal{O}_H\mathcal{O}_L]_{nl}\) are dual to a bulk field \(\phi_{nl}\). The mass of these modes written in terms of the energies of these modes \(\omega_{nl}\), \eqref{td}
\begin{equation}
    \Delta_{[HL]_{nl}}=\tau_n+|l|=\Delta_1 +\alpha(\Delta_2 +2n)+\omega_{nl}
\end{equation}
\section{Structure of the OPE}
There are two ways to understand how the heavy and light-particles interact in
the bulk. The first perspective is, the light state and the heavy state come together, and bind gravitationally to form a two-particle composite states. These composite states represent a light-particle orbiting around a heavy-particle. From another perspective, the light and heavy-states propagate independently and exchange gravitons and other virtual particles, to mediate the interactions between them. \\
In the holographic CFT, the binding and exchange mechanisms are analogous to the s- and t-channel expansions of the correlator. The heavy-light Virasoro block can be decomposed into the sum of primary and descendants three-point functions,
\begin{figure}[h]
    \centering
    \includegraphics[width=0.65\linewidth]{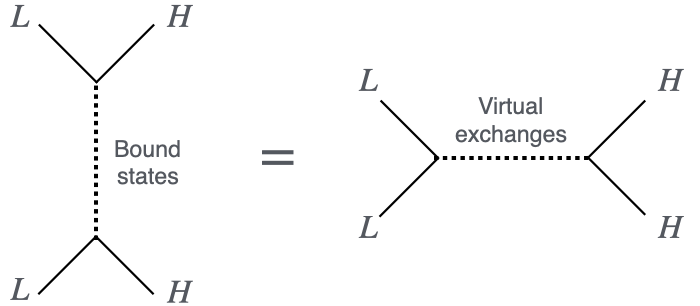}
    \caption{This figure illustrates the crossing for a heavy-light correlator.}
    \label{fig:enter-label}
\end{figure}
\begin{align}
    \mathcal{V}(z,\bar{z})=&\braket{\wick{\c1{\mathcal{O}_H} \c1{\mathcal{O}_L}\c1{\mathcal{O}_L}\c1{\mathcal{O}_H}}}=\sum_{\psi}\braket {\mathcal{O}_H\mathcal{O}_L|\psi}\braket{\psi|\mathcal{O}_H\mathcal{O}_L}=\sum_{h,\bar{h}}(P^{HL}_{h,\bar{h}})^2 \mathcal{F}^{s}_h\bar{\mathcal{F}}^{s}_{\bar{h}}
    \nonumber \\ =&\braket{\wick{\c1{\mathcal{O}_H}\c2{\mathcal{O}_L}\c2{\mathcal{O}_L}\c1{\mathcal{O}_H}}}=\sum_{\psi}\braket {\mathcal{O}_L\mathcal{O}_L|\psi}\braket{\psi|\mathcal{O}_H\mathcal{O}_H}=\sum_{h,\bar{h}}(P^{LL}_{h,\bar{h}})(P^{HH}_{h,\bar{h}}) \mathcal{F}^{t}_h\bar{\mathcal{F}}^{t}_{\bar{h}}
\end{align}
where the holomorphic and anti-holomorphic functions \(\mathcal{F}_h(z) \bar{\mathcal{F}}_{\bar{h}}(\bar{z})\) are the Virasoro blocks. The intermediate states that propagate in the OPE will constitute only the identity and the double-twist as \(\mathcal{O}_L\) is a generalised free field.
\begin{figure}[b]
    \centering
    \includegraphics[width=1.0\linewidth]{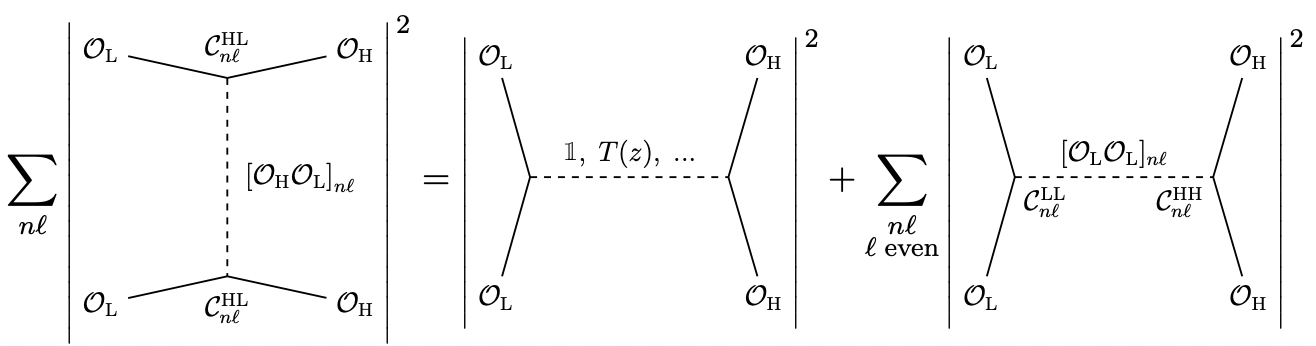}
    \caption{This figure displays the spectrum of the exchange states in the s-channel (left) and the t-channel (right)  \cite{grabovsky2024heavystates3dgravity}. }
    \label{fig:enter-label}
\end{figure}
In the s-channel \(\mathcal{O}_L\) and \(\mathcal{O}_H\) combine to form a HL double-twist operator \([\mathcal{O}_H\mathcal{O}_L]_{nl}\). The bulk picture is that  a light-mode \(\phi_{nl}=P^{HL}_{nl}e^{-i\omega_{nl}t}e^{il\theta}\) orbits a heavy object. \\ In the t-channel, pairwise identical operators are brought together. The leading order contribution is from the vacuum block, which encodes the gravitational interaction between \(\mathcal{O}_L\) and \(\mathcal{O}_H\). Furthermore, the sub-leading contribution comes from the double twist operators \([\mathcal{O}_L\mathcal{O}_L]_{nl}\) and \([\mathcal{O}_H\mathcal{O}_H]_{nl}\).
 \section{Twist accumulation near Black Holes}
 We will focus on determining the twist accumulation of the heavy-light bound states\([\mathcal{O}_H\mathcal{O}_L]_{n,l}\) in the BTZ black hole. Once gain, for the four-point function, but now the conformal dimension of the two primaries \(\mathcal{O}_1\) and \(\mathcal{O}_2\) are different,
 \begin{equation}
     \braket{\mathcal{O}_1\mathcal{O}_1\mathcal{O}_2\mathcal{O}_2}
 \end{equation}
 the conformal bootstrap equation is
 \begin{equation}
     \mathcal{V}(u,v)_{0,0}+\sum_{h,\bar{h}}P^{11,22}_{h,\bar{h}}\mathcal{V}^{11,22}_{h,\bar{h}}(u,v)=\left(\frac{u}{v}\right)^{\frac{\Delta_1+\Delta_2}{2}}u^{-\Delta_{12}}\sum_{h,\bar{h}}P^{12,12}_{h,\bar{h}}\mathcal{V}^{12,12}_{h,\bar{h}}(v,u)
 \end{equation}
 where \(\Delta_{12}=\Delta_1-\Delta_2\) and we have separated out the identity contribution in the s-channel. We know that the leading behaviour of the conformal block in small \(u\) and \(v\) 
 \begin{equation}
     \mathcal{V}(u,v)_{h,\bar{h}}=u^{\frac{\tau}{2}}\mathcal{F}_{h,\bar{h}}(u,v)
 \end{equation}
 where \(\mathcal{F}\) is an analytic function in \(u\). Using this, we can the identify block as 
 \begin{equation}
     \mathcal{V}(u,v)_{0,0}=\left(\frac{u}{v}\right)^{\frac{\Delta_1+\Delta_2}{2}}u^{-\Delta_{12}}\sum_{\tau,l}P^{12,12}_{\tau,l}\mathcal{F}_{\tau,l}(v,u) \;\ \text{when} \;\ u \rightarrow 0 \label{v0}
 \end{equation}
 We will evaluate the structure of this block in the heavy-light semi-classical limit with 
 \begin{equation}
     \frac{\Delta_1}{c} \ll 1 \;\ \text{and} \;\ \Delta_2=O(c)
 \end{equation}
 So the heavy-light identity block becomes,
 \begin{equation}
     \mathcal{V}(u,v)_{0,0}=\alpha^{\Delta_1}v^{-\frac{\Delta_1}{2}(1-\alpha)}\left(\frac{1-v}{1-v^{\alpha}}\right)^{\Delta_1} \label{hl}
 \end{equation}
 with \(\alpha=\sqrt{1-12\frac{\Delta_2}{c}}\). Using \eqref{hl} and \eqref{v0}, we can write 
 \begin{equation}
     1=\alpha^{-\Delta_1}v^{-\frac{1}{2}(\alpha\Delta_1+\Delta_2)}u^{\frac{\Delta_1+\Delta_2}{2}}u^{-\frac{1}{2}\Delta_{12}}\sum_{\tau,l}P^{12,12}_{\tau,l}\mathcal{F}_{\tau,l}(v,u)
 \end{equation}
 In the large \(l\) limit of the global blocks, we know that 
 \begin{equation}
      \mathcal{F}_{\tau,l}(v,u) \approx 2^{\tau+2l}v^{\frac{\tau}{2}}u^{\frac{\Delta_{12}}{2}}\sqrt{\frac{l}{\pi}}K_{\Delta_{12}}(2\sqrt{u})
 \end{equation}
 Inserting this into the bootstrap equation, and working in the light-cone bootstrap limit \((u \ll v \ll 1)\), we get
 \begin{equation}
     v^{\frac{\Delta_2+\alpha\Delta_1}{2}}\sum_{n=0}^{\infty} \frac{(\Delta_1)_n}{n!}v^{n\alpha}=\alpha^{-\Delta_1}u^{\frac{\Delta_1+\Delta_2}{2}}v^{\frac{\tau}{2}}\sum_{\tau,l}P_{\tau,l}2^{\tau+2l}\sqrt{\frac{l}{\pi}}K_{\Delta_{12}}(2\sqrt{u})
 \end{equation}
 Matching powers of \(v\) on both sides, we get that the twist accumulation occurs at 
 \begin{equation}
     \tau_n=\Delta_2+\alpha(\Delta_1+2n) \label{td}
 \end{equation}
 We find that the large \(l\) spectrum of any CFT with a large central charge \(c\), the twist of the heavy-light bound state ,matches with that of the light-light bound state after the following rescaling:
 \begin{equation}
     \Delta \rightarrow \Delta\sqrt{1-8G_N M}, \;\ n \rightarrow n\sqrt{1-8G_N M}
 \end{equation}
 This is the spectrum of the a heavy-light bound state orbiting a deficit angle in AdS\(_3\). \\ For the case of a BTZ black hole, \(\Delta_2 > \frac{c}{12}\), which implies that \(\alpha=\sqrt{1-12\frac{\Delta_2}{c}}=i\beta\). Making this change in \eqref{td}, we get 
 \begin{equation}
     \tau_n=\Delta_2+i\beta(\Delta_1+2n)=2i\pi T_H(\Delta_1+2n)+\Delta_2
 \end{equation}
 These are exactly the quasi-normal modes associated to a BTZ black hole, as in \cite{Birmingham_2002}.
\chapter{Witten Diagrams in AdS\(_3\)}
In the AdS/CFT correspondence, the role of conformal blocks has been lagging. The use of Witten diagrams in computing the spectral and OPE data of the dual CFT from a holographic correlation function has taken leaps forward since the introduction of Mellin space formalism \cite{Fitzpatrick_2011,Paulos_2011}. However, one sees that a systematic method of decomposing Witten diagrams into conformal blocks is missing. A natural question that seems to be unanswered is: namely, what object in AdS computes a conformal block? A geometric bulk description of a conformal block
would greatly aid in the comparison of correlators between AdS and CFT, and presumably allow for a more efficient implementation of the dual conformal block decomposition, as it
would remove the necessity of actually computing the full Witten diagram explicitly. There is a simple, yet elegant answer to this problem. The answer is that conformal blocks are computed by “geodesic Witten diagrams” \cite{Chen_2017**,Hijano_2016} The main feature of a geodesic Witten diagram that distinguishes it from a standard exchange Witten diagram is that in the former, the bulk vertices are not integrated over all of AdS, but only over geodesics connecting points on the boundary hosting the external operators. This representation of conformal blocks in terms of geodesic Witten diagrams is valid in all spacetime dimensions, and holds for all conformal blocks belonging to arbitrary CFTs. \\ In this section, we will be motivated to use the geodesic Witten diagrams to bulk construct Virasoro blocks \cite{Hijano_2015}. We will see how Witten diagrams reproduce the global block in the \(c \rightarrow \infty\) limit. Furthermore, in order to reproduce the heavy-light Virasoro blocks, we will allow one of the geodesics to back react on AdS\(_3\). This sets up a conical defect or a BTZ geometry for the remaining part of the geodesic Witten diagram corresponding to the light operators. Essentially to construct a heavy-light block, we will be using scalar fields propagating in the background of a locally AdS\(_3\) conical defect geometry.
\section{Global Blocks from AdS\(_3\) Gravity}
The global block simply means setting \(\alpha=1\) in \eqref{vira}. This is equivalent to holding all \(h_i\) fixed as \(c \rightarrow \infty\). Let's consider a global block corresponding to the exchange of a spinless operator \(\mathcal{O}_p\), such that \(h_p=\bar{h}_p \equiv \frac{\Delta}{2}\). In order to define a geodesic Witten diagram, we first define the an ordinary exchange Witten diagram in AdS\(_3\), with the exchanged scalar field having mass \(m^2=\Delta(\Delta-2)\). For the complete Witten diagram, the cubic
vertices should be integrated over all the AdS spacetime. However, for computing the geodesic Witten diagram, and as a consequence the global block, we restrict the integration to the bulk geodesics \(\gamma_{12}\) and \(\gamma_{34}\), connecting the boundary points on the Witten diagram. Then the geodesic Witten diagram for a scalar field exchange is,
\begin{equation}
    \mathcal{W}_{\Delta,0}(x_i)=\int_{\gamma_{12}} d\lambda\int_{\gamma_{34}} d\lambda^{'}G_{b\partial}(x_1,y(\lambda))G_{b\partial}(x_2,y(\lambda))G_{bb}(y(\lambda),y(\lambda^{'}),\Delta)G_{b\partial}(x_3,y(\lambda^{'}))G_{b\partial}(x_4,y(\lambda^{'})) \label{gw}
\end{equation}
\begin{figure}[h]
    \centering
    \includegraphics[width=0.35\linewidth]{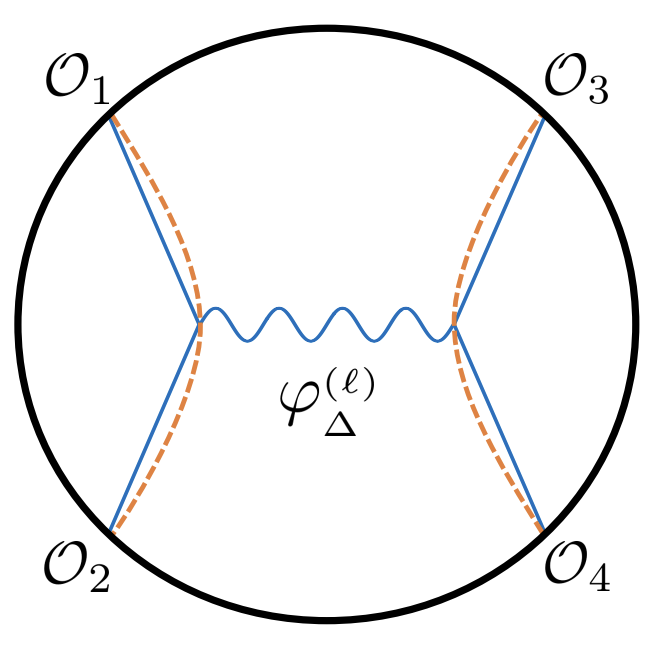}
    \caption{This illustrates a geodesic Witten diagram in AdS\(_{d+1}\) when the exchange state is a symmetric traceless spin \(l\) tensor having mass \(m^2=\Delta(\Delta-d)-l\). We integrate the vertices over the geodesics (dashed lines) connecting the two pairs of boundary points. This computes the Virasoro block for the exchange of a CFT\(_d\) operator having conformal weight \(\Delta\) and spin \(l\). In the case of dimension \(d=2\), the conformal block simply reduces to the product of the holomorphic and anti-holomorphic global blocks. In the case of the heavy-light Virasoro blocks, one of the geodesics back-react, creating a conical defect in the background geometry \cite{Hijano_2015}.}
    \label{fig:1}
\end{figure}
\\
 where \(\lambda\) and \(\lambda^{'}\) denote the proper length. Moreover \(G_{b\partial}\) and \(G_{bb}\) denote the bulk-to-boundary and bulk-to-bulk propagators. In our convention \(x\) denotes a point on the boundary, whereas \(y\) denotes a point in the bulk. The expression \eqref{gw} is equivalent to the product of the holomorphic and anti-holomorphic global blocks for an exchange of a spinless scalar field \(\mathcal{O}_p\).
\section{Virasoro heavy-light blocks in AdS\(_3\) Gravity}
We are now ready to deal with the case when \(\alpha \not= 0\). This is equivalent to taking the heavy-light limit instead of keeping all \(h_i\) fixed. This suggests that we scale up the dimensions \(h_1\) and \(h_2\) to \(O(c)\) for two conformal operators. This has a natural interpretation: let \(\gamma_{12}\) back react and we evaluate \eqref{gw} for this new geometry. The way this is done is by placing the heavy operators in the past and future infinity, then the geodesic \(\gamma_{12}\) back reacts on the AdS\(_3\) geometry to generate a conical defect or BTZ black hole geometry with the metric 
\begin{equation}
    ds^2=\frac{\alpha^2}{\cos^{2}\rho}\left(\frac{d\rho^2}{\alpha^2}+d\tau^2+\sin^{2}\rho d\phi^2 \right) \label{met}
\end{equation}
with \(\phi \cong \phi+2\pi\). In the case of \(\alpha < 1\), we have a conical defect with a singularity at \(\rho=0\). However, for \(\alpha^2<0\), we will have a BTZ black hole geometry after doing a Wick rotation. This geometry is generated from a particle source of mass \(m^2=4h_{H_1}(h_{H_1}-1)\) sitting in the origin of the global AdS with, 
\begin{equation}
    \alpha=\sqrt{1-24\frac{h_{H_1}}{c}}
\end{equation}
We are aware that the heavy-light block is simply
\begin{equation}
    \braket{\mathcal{O}_{H_1}(\infty,\infty)\mathcal{O}_{H_2}(0,0)[\mathcal{P}_h]\mathcal{O}_{L_1}(z,\bar{z})\mathcal{O}_{L_2}(0,0)}\rightarrow \mathcal{V}(h_i,h_p,c;z-1) \bar{\mathcal{V}}(\bar{h}_i,\bar{h}_p,c;\bar{z}-1) 
\end{equation}
where 
\begin{equation}
    \mathcal{V}(h_i,h_p,c;z-1)=z^{(\alpha-1)h_{L_1}}(1-z^{\alpha})^{h_p-h_{L_1}-h_{L_2}}\ _2F_1(h_p+h_{12},h_p-\frac{H_{12}}{\alpha},2h_p;1-z^{\alpha}) \label{hlb}
\end{equation}
with the definition \(h_{12} \equiv h_{L_1}-h_{L_2}\), \;\ \(H_{12} \equiv h_{H_1}-h_{H_2}\) and \(\omega=1-z^{\alpha}\). \\ It is convenient to re-write the heavy-light block in the cylindrical coordinate \(w=\phi+i\tau\). Under this coordinate transformation , we have 
\begin{equation}
    \mathcal{V}(h_i,h_p,c;w)=\sin^2\left(\frac{\alpha w}{2}\right)^{-2h_{L_1}}(1-e^{i\alpha w})^{h_p+h_{12}} \ _2F_1(h_p+h_{12},h_p-\frac{H_{12}}{\alpha},2h_p;1-e^{i\alpha w})
\end{equation}
Coming back to the task at hand, we need to compute \eqref{gw}, but now the propagators of the light operators in the conical defect metric \eqref{met} which is produced by the heavy operators. This can bethought of as the light particle geodesic and the conical defect exchanging a bulk field corresponding to the primary operator \(\mathcal{O}_p\). Once again, we will only consider spinless exchange operator with \(h=\bar{h}\). \\ We will now reproduce \eqref{hlb} using the geodesic Witten diagram in the conical defect background. 
\begin{align}
    &\mathcal{W}_{2h_p,0}(x_i)=\int_{-\infty}^{\infty} d\lambda\int_{-\infty}^{\infty} d\lambda^{'}G_{b\partial}(\tau_1=-\infty,\tau(\lambda))G_{b\partial}(\tau_2=-\infty,\tau(\lambda)) \nonumber \\  &\times \;\ G_{bb}^{\alpha}(y(\lambda),y(\lambda^{'}),2h_p)G_{b\partial}^{\alpha}(w_1=0,y(\lambda^{'}))G_{b\partial}^{\alpha}(w_2=w,y(\lambda^{'})) \label{whl}
\end{align}
where \(G^{\alpha}\) is the propagator in the conical defect metric and we have explicitly highlighted the \(\tau\)-dependence for the in- and out-states in the cylindrical picture. \\ Due to the \(O(d,2)\) isometry group of AdS, propagators can only depend on the invariant distance function \(\sigma\)
\begin{equation}
    \sigma=\frac{\cos(\tau-\tau^{'})-\sin{\rho_1}\sin{\rho_2}\cos(\phi-\phi^{'})}{\cos{\rho_1}\cos{\rho_2}}
\end{equation}
We define the geodesic distance \(\xi=\ln{(\sigma+\sqrt{\sigma^2-1})}\). In terms of the geodesic distance, the bulk-to-bulk propagator for global AdS is defined by 
\begin{equation}
    G_{bb}(y,y',2h)=\sigma^{-2h}\ _2F_1(h,h+\frac{1}{2},2h,\frac{1}{\sigma^2})=\frac{e^{2h\xi(y,y')}}{e^{2\xi(y,y')}-1}
\end{equation}
Furthermore, in the limit one of the fields approach the boundary , we get
\begin{equation}
    G_{b\partial}(x'y)=\left(\frac{\cos\rho}{\cosh{(\tau-\tau^{'})}-\sin\rho\cos{(\phi-\phi^{'})}}\right)^{2h}
\end{equation}
In order to evaluate these propagators in the conical defect geometry, we make the following replacements \(\tau \rightarrow \alpha\tau\) and \(\phi \rightarrow \alpha\phi\). However, there a subtlety here, as the periodicity \(\phi \approx \phi+2\pi\) is violated, the Virasoro block is not single-valued. However, this is not a problem as the Virasoro blocks are indeed multi-valued functions. These non single-valued propagators correctly reproduce the branch cuts of the Virasoro blocks. \\ The product of the heavy operator propagators, with end-points fixed at the past and future infinity, and evaluated at \(\rho=0\) is 
\begin{equation}
    G_{b\partial}(\tau_1=-\infty)G_{b\partial}(\tau_2=\infty)=e^{-2H_{12}\tau}
\end{equation}
We should notice that \(\lambda=\alpha\tau\) denotes the proper time. To define a bulk-to-boundary propagator for the light internal and external operators in the conical defect, we consider a geodesic connecting two points \(w_1 and w_2\) on the boundary. The geodesic begins at \(w_1\) and ends at \(w_2\) and they lie on the same time slice, so that \(w_{12}=w_1-w_2\) is a real quantity. We then define 
\begin{equation}
    \cos\rho(\lambda)=\frac{\sin{\frac{\alpha w_{12}}{2}}}{\cosh{\lambda}} \;\ , \;\ e^{2i\alpha w(\lambda)}=\frac{\cosh({\lambda}-\frac{i\alpha w_{12}}{2})}{\cosh({\lambda}+\frac{i\alpha w_{12}}{2})}e^{i\alpha(w_1+w_2)}
\end{equation}
The bulk-to-boundary propagator for the light-fields, evaluated on these geodesics become
\begin{align}
   &G_{b\partial}^{\alpha}(w_1,y(\lambda^{'}))=\frac{e^{-2\lambda^{'}h_{L_1}}}{(\sin\left({\frac{\alpha w_{12}}{2}}\right))^{2h_{L_1}}} \nonumber \\ &G_{b\partial}^{\alpha}(w_2,y(\lambda^{'}))=\frac{e^{2\lambda^{'}h_{L_2}}}{(\sin\left({\frac{\alpha w_{12}}{2}}\right))^{2h_{L_1}}}
\end{align}
We now move onto the bulk-to-bulk propagator with the exchange field of dimension \(h_p\) evaluated with one endpoint at \(\rho=0\) at time \(\tau\) and the other endpoint is held fixed at time \(\tau^{'}=0\),
\begin{equation}
  G_{bb}(y,y',2h)=\sigma^{-2h}\ _2F_1(h_p,h_p+\frac{1}{2},2h_p;\frac{1}{\sigma^2}) \;\ , \;\ \sigma=\frac{\cosh{\lambda}\cosh{\lambda^{'}}}{\sin\left({\frac{\alpha w_{12}}{2}}\right)}   
\end{equation}
Plugging these values into \eqref{whl} and setting \(w_1=0\) and \(w_2=w\), we get the following integral expression
\begin{align}
    &\mathcal{W}_{2h_p,0}(w)=\left(\sin\left({\frac{\alpha w_{12}}{2}}\right)\right)^{2h_p-2h_{L_1}-2h_{L_2}}\int_{-\infty}^{\infty} d\lambda\int_{-\infty}^{\infty} d\lambda^{'} e^{-\frac{2H_{12}}{\alpha}\lambda-2h_{12}\lambda^{'}}(\cosh{\lambda}\cosh{\lambda^{'}})^{-2h_p} \nonumber \\ &\times \;\ _2F_1\left(h_p,h_p+\frac{1}{2},2h_p;\frac{\sin\left({\frac{\alpha w_{12}}{2}}\right)}{\cosh{\lambda}\cosh{\lambda^{'}}}\right)
\end{align}
The integral can be evaluated in the following manner: \\ We define the following quantity,
\begin{equation}
    \mathcal{W}_{2h_p,0}(w)=\left(\sin\left({\frac{\alpha w_{12}}{2}}\right)\right)^{2h_p-2h_{L_1}-2h_{L_2}} \times \mathcal{I}
\end{equation}
where,
\begin{align}
     &\mathcal{I}=\int_{-\infty}^{\infty} d\lambda e^{-\frac{2H_{12}}{\alpha}}(\cosh{\lambda})^{-2h_p} \nonumber \\ &\times \;\ \int_{-\infty}^{\infty} d\lambda^{'} e^{-2h_{12}\lambda^{'}}(\cosh{\lambda^{'}})^{-2h_p}\ _2F_1\left(h_p,h_p+\frac{1}{2},2h_p;\frac{\sin^2\left({\frac{\alpha w_{12}}{2}}\right)}{\cosh{\lambda}\cosh{\lambda^{'}}}\right)
\end{align}
The integral has a divergent behaviour in the regimes of large \(\lambda\) and \(\lambda^{'}\),
\begin{equation}
    |\frac{H_{12}}{\alpha}|<h_p \;\ \text{and} \;\  |h_{12}|< h_p \label{con}
\end{equation}
Writing the hypergeometric function as a series, we find 
\begin{align}
    \mathcal{I}=&\sum_{n=0}^{\infty} \left(\int_{-\infty}^{\infty} d\lambda \ e^{-\frac{2H_{12}}{\alpha}\lambda}(\cosh{\lambda})^{-2h_p-2n}\right) \left(\int_{-\infty}^{\infty} d\lambda^{'} e^{-2h_{12}\lambda^{'}}(\cosh{\lambda^{'}})^{-2h_p-2n}\right)\nonumber \\ & \times \frac{(h_p)_n (h_p+\frac{1}{2})_n}{(2h_p)_n n!}\left(\sin^2\left({\frac{\alpha w_{12}}{2}}\right)\right)^n \label{I}
\end{align}
We use the following identity of the Beta function \(B(a,b)\)
\begin{equation}
    \int_0^{\infty}dt \frac{\cosh{2bt}}{(\cosh{t})^{2a}}=4^{a-1}B(a+b,a-b) \;\ \text{for}\;\ \mathrm{Re}(a)>\mathrm{Re}(b) 
\end{equation}
The integrals are finite due to the constraints \eqref{con}, thus the Beta function identity can be manipulated to the following form 
\begin{align}
    &\int_{-\infty}^{\infty} d\lambda e^{-\frac{2H_{12}}{\alpha}\lambda}(\cosh{\lambda})^{-2h_p-2n} = 2^{2m-1} B(m-\frac{H_{12}}{\alpha},m+\frac{H_{12}}{\alpha}) \nonumber \\ &\int_{-\infty}^{\infty} d\lambda^{'} e^{-2h_{12}\lambda^{'}}(\cosh{\lambda^{'}})^{-2h_p-2n} = 2^{2m-1} B(m-h_{12},m+h_{12})
\end{align}
where \(m=n+h_p\). We substitute for the integrals in \eqref{I} and use the Legendre duplication formula to simplify the expressions,
\begin{equation}
    \Gamma(2t)=\frac{2^{2t-1}}{\sqrt{\pi}}\Gamma(t)\Gamma(t+\frac{1}{2})
\end{equation}
In our case, we get the following expression
\begin{equation}
    \Gamma(2h_p+2n)=2^{2n}\Gamma(2h_p)(h_p)_n (h_p+\frac{1}{2})_n
\end{equation}
The integral \(\mathcal{I}\) now becomes,
\begin{align}
    &\mathcal{I}=\frac{2^{4h_p-2}\Gamma(h_p+\frac{H_{12}}{\alpha})\Gamma(h_p-\frac{H_{12}}{\alpha})\Gamma(h_p+h_{12})\Gamma(h_p-h_{12}))}{\Gamma(2h_p)\Gamma(2h_p)} \nonumber \\ &\times \sum_{n=0}^{\infty} \frac{(h_p+\frac{H_{12}}{\alpha})_n (h_p-\frac{H_{12}}{\alpha})_n (h_p-h_{12})_n (h_p+h_{12})_n}{(2h_p)_n (h_p)_n (h_p+\frac{1}{2})_n n! } \left(\sin^2\left({\frac{\alpha w_{12}}{2}}\right)\right)^n \label{Inew}
\end{align}
We note, that the last line of the above equation can be written in terms of a generalised hypergeometric function  
\begin{equation}
    \pFq{p}{q}{a_1,...,a_p}{b_1,...,b_q}{z}=\sum_{k=0}^{\infty}\frac{(a_1)_k....(a_p)_k}{(b_1)_k....(b_q)_k}\frac{z^k}{k!}
\end{equation}
Then the sum in the last line becomes a power series in \(_4F_3\) hypergeometric function. The geodesic Witten diagram for heavy-light fields becomes 
\begin{equation}
   \mathcal{W}_{2h_p,0}(w)=\mathcal{N}\left(\sin{\frac{\alpha w_{12}}{2}}\right)^{2h_p-2h_{L_1}-2h_{L_2}}\pFq{4}{3}{h_p+\frac{H_{12}}{\alpha},h_p-\frac{H_{12}}{\alpha},h_p+h_{12},h_p-h_{12}}{2h_p,h_p,h_p+\frac{1}{2}}{\sin^2\left({\frac{\alpha w_{12}}{2}}\right)}
\end{equation}
where the \(\mathcal{N}\) stands for the overall numerical factor coming from the first line of \eqref{Inew}. It will be convenient for our to write the \(_4F_3\) hypergeometric function as a product of two \(_2F_1\) functions. In order to facilitate this, we employ the identity 
\begin{align}
    &\pFq{4}{3}{a,b,c,a+b-c}{\frac{a+b}{2},\frac{a+b+1}{2},a+b}{\frac{z^2}{4(z-1)}}=(1-z)^a\pFq{2}{1}{a,c}{a+b}{z}\pFq{2}{1}{a,a+b-c}{a+b}{z} \;\ , z \in (1,\infty) 
\end{align}
and 
\begin{align}
    &(1-z)^{a'}\pFq{2}{1}{a',b'}{c'}{z}=\pFq{2}{1}{a',c'-b'}{c'}{\frac{z}{z-1}}
\end{align}
We then get 
\begin{equation}
    \pFq{4}{3}{a,b-a,c,b-c}{\frac{b}{2},\frac{b+1}{2},b}{\frac{z^2}{4(z-1)}}=\pFq{2}{1}{a,c}{b}{z} \pFq{2}{1}{a,c}{b}{\frac{z}{z-1}}
\end{equation}
where 
\begin{equation}
    z=1-e^{i\alpha w_{12}}, \;\ a=h_p+h_{12}, \;\ c=h_p-\frac{H_{12}}{\alpha}, \;\ b=2h_p
\end{equation}
After the end of all these simplifications, once can see that the heavy-light geodesic Witten diagram expression boils down to
\begin{align}
     &\mathcal{W}_{2h_p,0}(w)=\mathcal{N}\left(\sin\left({\frac{\alpha w_{12}}{2}}\right)\right)^{2h_p-2h_{L_1}-2h_{L_2}} \times \pFq{2}{1}{h_p+h_{12},h_p-\frac{H_{12}}{\alpha}}{2h_p}{1-e^{i\alpha w_{12}}} \nonumber \\ &\times \pFq{2}{1}{h_p+h_{12},h_p-\frac{H_{12}}{\alpha}}{2h_p}{1-e^{-i\alpha w_{12}}}
\end{align}
This matches with the result evaluated using the heavy-light Virasoro blocks \eqref{hlb}. To summarize this section, we introduced a simple bulk prescription for reproducing the semi-classical heavy-light Virasoro blocks, involving a light-particle geodesic interacting with a heavy-particle defect via a light-particle exchange. \\ There are avenues of research, that which have not yet been explored to the full extent. 
\begin{itemize}
    \item \textbf{Loop calculations}: There hasn't been much success in calculating geodesic Witten diagrams to general loop-orders. However, there are some special loop diagrams, that have been calculated using this technique as demonstrated in Figure \ref{fig:loop diag}. It is only for such loop diagrams that the Mellin amplitudes are known. \cite{Penedones_2011}. It will interesting to see the \(1/c\) corrections to the global and Virasoro conformal block and how that's translated to the loop graviton calculations.
    \begin{figure}[h]
        \centering
        \includegraphics[width=0.70\linewidth]{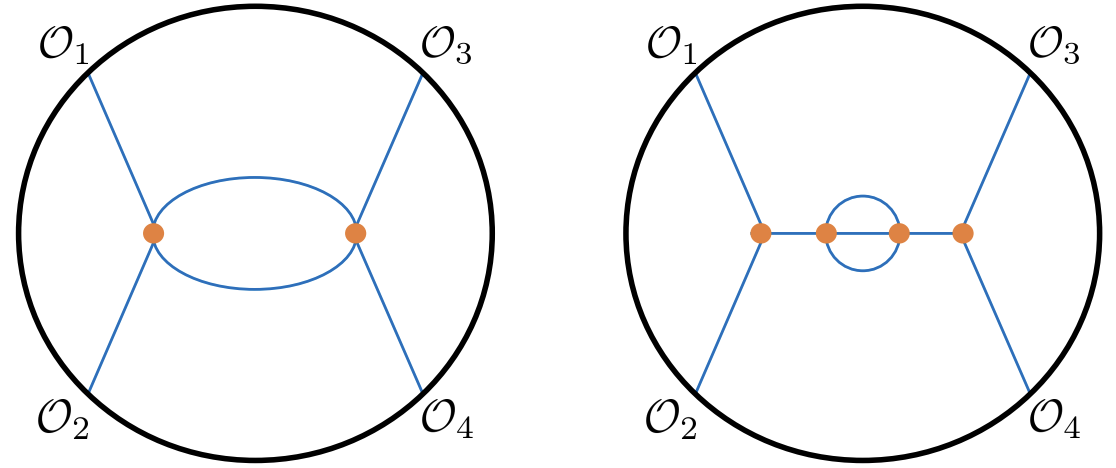}
        \caption{This figure depicts a special class of loop diagrams that can be represented as an infinite sum over tree-level exchange diagrams \cite{Hijano_2016}.}
        \label{fig:loop diag}
    \end{figure}
    \item \textbf{External operators with higher spin}: W-symmetry is an extension of 2d CFTs, with W-algebra being a higher-spin variant of the usual Virasoro algebra. It's natural to extend this analysis to heavy-light blocks with W-symmetry. Although, semi-classically the W\(_N\) conformal blocks for vacuum exchange have already been computed in \cite{deBoer:2014sna} with charges of the order \(c\).
    \item \textbf{Heavy-Heavy blocks}: We are yet to find a closed-form expression of a semi-classical Virasoro block with all heavy operators, as it is the limit of the Liouville theory. The general expectation that bulk dual is a spacetime interacting with a conical defect. It plays a key role in the study of two-interval Rènyi entropies \cite{hartman2013entanglemententropylargecentral,faulkner2013entanglementrenyientropiesdisjoint}.
\end{itemize}
\chapter{Information Loss in Black Holes}
\begin{figure}[t]
    \centering
    \includegraphics[width=0.70\linewidth]{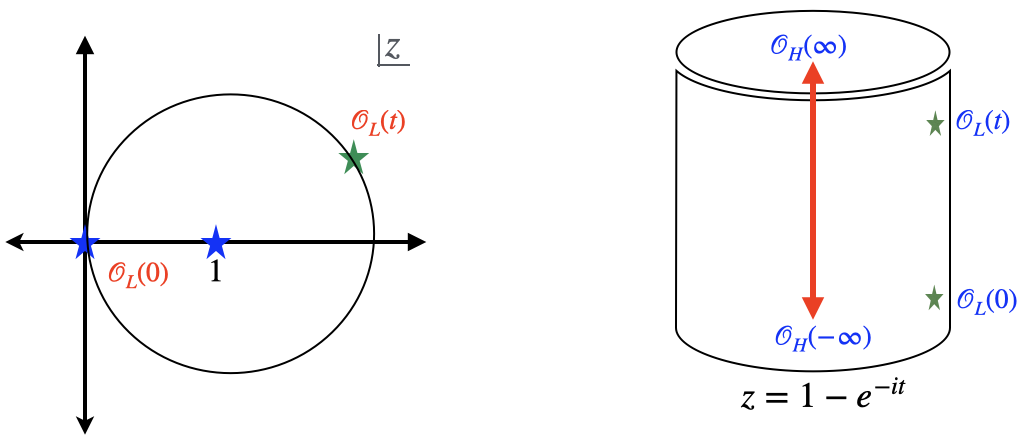}
    \caption{This figure depicts a Lorentzian heavy-light correlator. As time \(t\) increases in the cylindrical picture, the operator \(\mathcal{O}_L(t)\) periodically passes through the future lightcone of the operator \(\mathcal{O}_L(0)\). The Lorentzian correlator is given by  \(\braket{\mathcal{O}_H(\infty)\mathcal{O}_L(t_1)\mathcal{O}_L(t_2)\mathcal{O}_H(-\infty)}\), where \(t_1-t_2\) denotes the Lorentzian time separation in the CFT correlator.}
    \label{fig:cylinder}
\end{figure}

Black hole thermodynamics can be derived as a universal consequence of the Virasoro symmetry algebra of AdS\(_3\)/CFT\(_2\) \cite{Fitzpatrick_2014,Fitzpatrick_2015,Fitzpatrick_2016,Alkalaev_2015,Alkalaev_2016,Hijano_2015,Hartman_2014,Beccaria_2016,Besken_2016,fitzpatrick2016conformalblockssemiclassicallimit}. Unitarity is violated, i.e. the phenomena of Information Loss occurs in the semi-classical limit where Newton’s constant \(G_N \rightarrow 0\). The information loss effects persists order-by-order in a perturbation expansion in \(G_N=\frac{3}{2c}\). However, we will later see that the information loss can be resolved by non-perturbative effects \(e^{-c}\) \cite{fitzpatrick2016informationlossads3cft2,Chen_2017*}. The information loss problem are of two varieties. The \enquote{hard} information loss is the discord between local gravitational EFT and the unitary quantum mechanics. According to the AdS/CFT point of view, unitarity dominates, however doesn't explain how the Equivalence principle holds. To address this issue we will require a self-consistent prescription for reconstructing local bulk observable near and across horizons using only CFT data. This problem is extremely non-trivial to formulate in terms of quantum mechanical observables in CFT. However, there also exists an \enquote{easy} information loss paradox that can expressed in terms of CFT correlation functions. A two-point correlator which is probing a large AdS black hole will decay exponentially at late Lorentzian times. This implies that all information pertaining to an object thrown into a black hole will eventually be lost, signaling a violation of unitarity. A two-point thermal correlator of a scalar primary defined on a cylinder
\begin{equation}
    \braket{\mathcal{O}(t_L)\mathcal{O}(0)}_{T_H}=\left(\frac{\pi T_H}{\sinh{(\pi T_{H}t_L)}}\right)^{2\Delta_{\mathcal{O}}} \label{corr}
\end{equation}
decays exponentially at late Lorentzian times \(t_L\). The \enquote{easy} information paradox can be linked to the \textit{thermal} nature of black holes. To resolve this, we will need to identify and understand the non-perturbative \(e^{-\frac{1}{G_N}}\) corrections, which help restore unitarity. \\ A thermal two-point correlator of two light operators can be written as four-point correlation function,
\begin{equation}
    \braket{\mathcal{O}_L(1)\mathcal{O}_L(z,\bar{z})}_{T_H}\approx \braket{\mathcal{O}_H(\infty)\mathcal{O}_L(1)\mathcal{O}_L(z,\bar{z})\mathcal{O}_H(0)}
\end{equation}
where \(z\) describes the coordinate of a plane, while the Euclidean time \(t_E \equiv -\log(1-z)\) is defined on the boundary of the global AdS cylinder. This conformal transformation is depicted in Figure \ref{fig:cylinder}.
The heavy operator \(\mathcal{O}_H\) creates a black hole geometry with Hawking temperature \(T_H\). If the two-point correlator is thermal, it must satisfy the KMS condition, i.e. imposing a periodicity in Eucliean time \(t_E \sim t_E+\beta\). This \textit{periodic} behaviour is \textit{forbidden} in the vacuum correlators. In a Euclidean CFT, correlators can only have OPE singularities, when the operators collide. However, this periodicity in the Euclidean time gives rise to additional singularities at the periodic images of the OPE singularity such as \(z=\bar{z}=1-e^{\frac{n}{T_H}}\) for any integer \(n\). These \enquote{forbidden singularities} are the manifestation of unitarity violation and information loss. \\ In general, we expect that vacuum block’s information loss must be resolved within its own structure. The reasoning behind it is as follows: at the positions of these forbidden singularities, \(z=1-e^{\frac{n}{T_H}}\) is real and positive for \(n<0\) as well as \(T_H\) is real, thus the sum over conformal blocks is a sum over positive contributions. It implies that the sum over non-vacuum blocks cannot cancel this singular behavior \footnote{In the semi-classical limit \(c \rightarrow \infty\), the Vacuum block's singularities are sharper than those of the other Virasoro blocks.}. A more detailed proof states that the vacuum block can be viewed as an inner product between normalizable states, and OPE singularities can only exist at finite central charge \cite{Pappadopulo_2012}.
\section{Forbidden Singularities in AdS}
\begin{figure}
    \centering
    \includegraphics[width=0.95\linewidth]{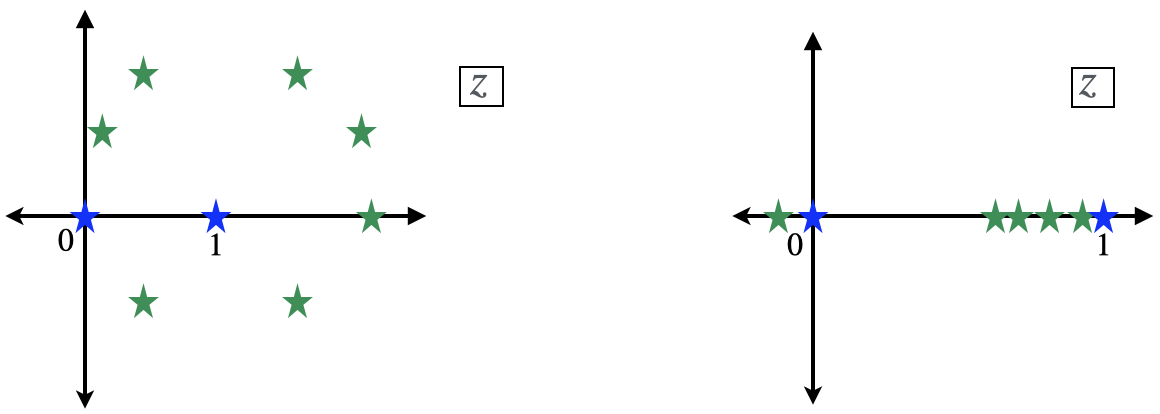}
    \caption{This figure shows the positions of the singularities that are forbidden in the unitary four-point correlators. The locations of the heavy operators in the Euclidean plane are 1 and \(\infty\), while the light operators are at 0 and \(z\).}
    \label{fig:sing}
\end{figure}

In the global coordinates of AdS spacetime, with the radius of curvature being  \(R_{AdS}=1\). Then the pure AdS metric becomes,
\begin{equation}
    ds^2=-(r^2+1)dt_L^2+\frac{dr^2}{(r^2+1)}+r^2 d\Omega^2
\end{equation}
which corresponds to a CFT on a cylinder \(R \times S^{d-1}\). In the thermal AdS phase, we compactify the Euclidean time \(t_E \sim t_E+\beta\). The Euclidean time periodicity implies that for \(\Omega=0\), there is an OPE singularity at \((t_E=\pm \beta,\pm 2\beta..)\) as a consequence of the geometry. Transforming the CFT from a cylinder to the plane via the coordinate transformation \(z=1-e^{-t+i\phi}\), these singularities occurs in the Euclidean regime \(z=\bar{z}=1-e^{\frac{n}{T_H}}\) for any integer \(n\). We see that AdS\(_3\) correlators for heavy operators and for operators above the BTZ threshold, these correlators have OPE image singularities. 
\\ The Euclidean metric of the BTZ black hole or the deficit angles is 
\begin{equation}
    ds^2=-(r^2+r_+^2)dt_L^2+\frac{dr^2}{(r^2+r_+^2)}+r^2 d\Omega^2 \label{metric}
\end{equation}
where the horizon radius is related to the Hawking temperature by \(r_+=2\pi T_H\). The BTZ black hole is dual to a thermal state in CFT of conformal dimension \(h_H\) and is related to the horizon radius via \(r_+=\sqrt{\frac{24h_H}{c}-1}\). Deficit angles are obtained by analytically continuing to imaginary \(r_+\), which happens when \(h_H < \frac{c}{24}\). \\ We have earlier stated, the deficit angles and BTZ black holes are orbifolds of AdS\(_3\) \cite{Ba_ados_1992}. This allows us to obtain the correlator \eqref{corr} using the method of images \cite{Rychkov_2017},
\begin{equation}
    \braket{\mathcal{O}_L(1)\mathcal{O}_L(z,\bar{z})}_{r_+}=\sum_{n=-\infty}^{\infty}[\mathcal{V}(z,n)]^{h_L}[\bar{\mathcal{V}}(\bar{z},n)]^{\bar{h}_L} \label{abcd}
\end{equation}
where 
\begin{equation}
    \mathcal{V}(z,n)=\frac{(1-z)}{\sin^2{\left(\frac{r_+}{2}\log(1-z)+2\pi in \right)}} \label{V}
\end{equation}
The sum over \(n\) ensures that the correlator is single-valued in a Euclidean plane and the singularities lie on a single Riemann sheet. So the imaginary part of \(\log(1-z)\) varies from 0 to \(2\pi i\), and for the real \(r_+\) the only non-singular term is the case when \(n=0\), whereas the singular term is 
\begin{equation}
    z=1-e^{\frac{2\pi m}{r_+}}
\end{equation}
for all integer \(m\), and this includes \(z=0\) for \(m=0\). For real \(r_+\)  these singularities lie on the real axis, while for imaginary \(r_+\) they form a unit circle around \(z=1\) , as shown in Figure \ref{fig:sing}. They are the periodic images of the OPE singularity arising from \(\mathcal{O}_{L}(z)\mathcal{O}_{L}(0)\sim\frac{1}{z^{2h_L}}\)
\section{Forbidden singularities in the Vacuum Block}
An noteworthy feature of the AdS\(_3\) correlator \eqref{abcd} is that for a real \(r_+\), the forbidden singularities contribute only in the \(n=0\) term of the image sum. This feature has an important interpretation the it's dual CFT. We can interpret \eqref{abcd} as a heavy-light correlator in CFT. The general structure of a four-point correlator in CFT\(_2\) is as follows,
\begin{equation}
    \braket{\mathcal{O}_H(\infty)\mathcal{O}_L(1)\mathcal{O}_L(z)\mathcal{O}_H(0)}=\mathcal{V}_0(1-z)\mathcal{V}_0(1-\bar{z})+\sum_{h,\bar{h}}P_{h,\bar{h}}\mathcal{V}_h(1-z)\mathcal{V}_{\bar{h}}(1-\bar{z})
\end{equation}
The factor \(\mathcal{V}_0\) is the identity part of the vacuum block. The vacuum block matches perfectly with the \(n=0\) term in the AdS sum image. \\ It has been shown in \cite{Fitzpatrick_2014} that \(\mathcal{V}(z,n)\) in \eqref{V} can be obtained from a bulk computation and is identical to the vacuum  block \(\mathcal{V}_0(1-z)\) in the limit \(c \rightarrow \infty\) with \(h_L\) and \(h_H/c\) fixed. This means that in the \(c\) limit, heavy-light CFT correlators have forbidden singularities that must be resolved at finite \(c\). Furthermore, at finite \(c\) the singularities must be resolved within the structure of the vacuum block itself. This is true because the vacuum block always produces the most singular term \(\propto z^{-2h_L}\) in the OPE limit \(z \rightarrow 0\). The other conformal blocks behave like \(\propto z^{h-2h_L}\) in the OPE singularity limit, with \(h>0\) for unitary theories and \(h=0\) for conserved currents. Since the forbidden singularities are images of the OPE singularities, other conformal blocks will always be less singular than the vacuum block in both the OPE and image singularity limits. To see this fact explicitly, we start out with a heavy-light block
\begin{equation}
    \mathcal{V}_{h_I}(z) \propto \left(\frac{1-w}{1-z}\right)^{h_I} w^{h_I-2h_L} \ _2F_1{(h_I,h_I,2h_I;w)}
\end{equation}
where \(w=1-(1-z)^{ir_+}\) and \(r_+=\sqrt{\frac{24h_H}{c}-1}\). The strength of the forbidden singularity is reduced when the conformal weight of the exchange operator \(h_I>0\). When \(T_H\) is real, the forbidden singularity at \(z=1-e^{\frac{n}{T_H}}\) for \(n<0\) are located at real and positive \(z\), and therefore the sum over the other conformal blocks converges over positive contributions that only add to the singularity in the vacuum block and not cancel it.
\section{Correlators at Late Lorentzian Times}
The research of Maldacena \cite{Maldacena_2003} emphasized that in a black hole background, correlators decay exponentially at late Lorentzian times. The consequence is that a small perturbation to the initial density matrix becomes scrambled at late times \cite{Bak_2007,Anous_2016,Germani_2014}. This means that information thrown into a black hole is lost. Such behavior is forbidden in a theory which has a finite number of local degrees of freedom on a compact space. This is a strong indicator of information loss when CFT correlators are derived from AdS. \\ We show that heavy-light Virasoro blocks in the limit \(h_I\) fixed and \(h_H>\frac{c}{24}\) with \(c \rightarrow \infty\) vanish exponentially when analytically continued to late Lorentzian times. To verify this claim, we again start of with a heavy-light Virasoro block,
\begin{equation}
    \mathcal{V}_{h_I}(z) \propto \left(\frac{1-w}{1-z}\right)^{h_I} w^{h_I-2h_L} \ _2F_1{(h_I,h_I,2h_I;w)} \;\ ,\;\ w=1-(1-z)^{ir_+}
\end{equation}
with \(r_+=2\pi T_H=\sqrt{\frac{24h_H}{c}-1}\) and \(h_H>\frac{c}{24}\) corresponding to a BTZ black hole in AdS\(_3\). Moreover, we know \(z=1-e^{-it_L}\) and since \(\alpha\) is imaginary, we have \(w=1-e^{2\pi T_{H} t_L}\). The Virasoro block at large \(t_L\) can be approximated,
\begin{equation}
   _2F_1\left(h_I,h_I,2h_I,1-e^{2\pi T_{H} t_L}\right) \propto e^{-2\pi h_I T_{H} t_L}
\end{equation}
So the heavy-light Virasoro at late Lorentzian times \(t_L \rightarrow \pm \infty\) are \(\propto e^{-2\pi h_I T_{H} t_L}\). Thus all heavy-light, large central charge Virasoro blocks vanish at large Lorentzian times. As we expect the sum over the conformal blocks to be convergent in CFT\(_2\), this implies that the correlators constructed from this sum must also vanish exponentially at late Lorentzian times \(t_L\). There exists a open question at this moment: Currently we do not have explicit expression for Virasoro blocks when the we have heavy exchange operators \(h_I \sim O(c)\), thus there always lies the possibility that such Virasoro blocks do not decay exponentially at late \(t_L\). \\ What we will be interested in is understanding how \textit{exact} Virasoro blocks with \(h_H>\frac{c}{24}\) behave at late Lorentzian time limit. We will address this version of information loss in the subsequent sections and will see that the behaviour of the vacuum block changes qualitatively around the (Page) time, which is of order \(S=\frac{\pi^2}{3}cT_H\), the black hole entropy.
\chapter{Exact Virasoro Blocks}
An ideal tool to tackle the problem of information loss would have been a explicit closed-form expression for general Virasoro blocks. Such an expression would have allowed us to observe how the forbidden singularities and late Lorentzian time behaviour are resolved by non-perturbative \(e^{-c}\) effects in the large \(c\) expansion. Present tools provide recursion relations \cite{Zamolodchikov1984} that help compute the series expansion \cite{Perlmutter2015} of the blocks near \(z=0\) with generic \(h_i,h,c\). The heavy-light blocks displays forbidden singularities at large \(c\). However, none of these provide an hint on how these singularities are resolved by finite \(c\) effects. Additionally, the relation between general semi-classical Virasoro blocks to the Painlevé \RNum{6} \cite{Litvinov_2014}, which can be solved in terms of it's own special function, does not give us any hope of writing down a closed-form expression for the Virasoro block \(\mathcal{V}\). \\ However, in CFT\(_2\) there exists a class of exact Virasoro blocks with special values of \(h_i,h,c\) \cite{DiFrancesco:1997nk, ginsparg1988appliedconformalfieldtheory,BENOIT1988517,BELAVIN1984333}. We start by defining the highest weight state \(\ket{h}=\mathcal{O}(0)\ket{0}\), which satisfies \(L_{0}\ket{h}=h\ket{h}\) and is created by the action of a primary \(\mathcal{O}\) acting on the sl(2,R) invariant vacuum \(\ket{0}\), which satisfies \(L_{n}\ket{0}=0, \ n \geq -1\). Descendent states are created by acting on \(\ket{h}\) with a string of \(L_{-n_i}\)'s, \(n_i>0\). Starting from the highest weight state \(\ket{h}\), we build the set of states,
\begin{center}
    \begin{tabular}{c|c|c}
      level   &  dimension & state \\ \hline
        0 & h & \(\ket{h}\) \\ 
        1 & h+1 & \(L_{-1}\ket{h}\)\\ 
        2 & h+2 & \(L^2_{-1}\ket{h},L_{-2}\ket{h}\)\\ 
        3 & h+3 & \(L^3_{-1}\ket{h},L_{-2}L_{-1}\ket{h},L_{-3}\ket{h}\)\\
        & ... & \\ 
        N & h+N & P(N) states
    \end{tabular}
\end{center}
known as Verma module. A linear combination of states that vanishes is known as a null state, and the representation of the Virasoro algebra with highest weight \(\ket{h}\) is constructed from the above Verma module by removing all null states (and their descendants). These CFTs which contain a finite number of irreducible representations of the Virasoro algebra are called Minimal models. It turns out that these Virasoro minimal models are extremely useful in understanding physical systems during their second order phase transitions. Explicitly, the minimal models describe the scaling limit of critical lattice models in statistical physics. For instance, according to \cite{Friedan:1983xq}, the scaling limit of the Ising model, tricritical Ising model, 3-state Potts model, and tricritical 3-state Potts model are described respectively by the main series minimal models \(\mathcal{M}(3,4),\mathcal{M}(4,5),\mathcal{M}(5,6),\) and \(\mathcal{M}(6,7)\). There are various other equally important applications of minimal models, presently which we won't delve into. \\ In the case of these exact conformal blocks, the external operators are degenerate, i.e. some of the Virasoro descendants are null states, which means that the Verma module is reducible. For describing degenerate blocks, it is convenient to define a parameter \(b\),
\begin{equation}
    c=1+\left(b+\frac{1}{b}\right)^2
\end{equation}
when we take the semi-classical limit \(c \rightarrow \infty\), we can take either \(b \rightarrow 0\) or \(b \rightarrow \infty\). The simplest reducible Verma module at level 2 is
\begin{equation}
    (L^2_{-1}+b^{2}L_{-2})\ket{h_{1,2}}=0 \label{diff}
\end{equation}
The \textit{Gram} matrix for level 2 Verma module is defined as follows,
\begin{align}
   M^{(2)}= \begin{pmatrix}
    \braket{h|L^2_{1}L^2_{-1}|h} & \braket{h|L^2_{1}L_{-2}|h} \\
    \braket{h|L_{2}L^2_{-1}|h} & \braket{h|L_{2}L_{-2}|h} \label{matrix}
\end{pmatrix}
\end{align}
The representation is not unitary if the determinant or the trace of the \textit{Gram} matrix is negative. The vanishing determinant of the \textit{Gram} matrix determines the number of null states at any given level of the \(N\). For instance for the level \(N=2\), 
\begin{align}
    \det M^{(2)}=\det\begin{pmatrix}
      4h+\frac{c}{2} & 6h \\
      6h & 4h(2h+1)
\end{pmatrix}
=32(h-h_{1,1})(h-h_{1,2})(h-h_{2,1})
 \end{align}
where the roots of \(\det M^{(2)}(c,h)\) are
\begin{align}
    &h_{1,1}=0 \nonumber \\ &h_{1,2}=\frac{1}{16}\left(5-c - \sqrt{(1-c)(25-c)}\right) \nonumber \\ &h_{2,1}=\frac{1}{16}\left(5-c + \sqrt{(1-c)(25-c)}\right)
\end{align}
and \(h_{1,1}=0\) root is due to the null state at \(L_{-1}\ket{0}=0\). Generally, degenerate states can only have special values of the conformal weight given by 
\begin{equation}
    h_{r,s}(b)=\frac{b^2(1-r^2)}{4}+\frac{(1-s^2)}{4b^2}+\frac{(1-rs)}{2}
\end{equation}
for positive integers \(r\) and \(s\). This formula determines the conformal weight of the null state when the \textit{Kac} determinant \eqref{matrix} is vanishes. Another interesting feature of this equation is that it has a symmetry, \(r \leftrightarrow s\) simply corresponds to \(b \leftrightarrow 1/b\). \\ The relation \eqref{diff} becomes an extremely useful differential equation for correlation functions of primary operators \(\mathcal{O}_{1,2}\) which produce states of conformal weight \(h_{1,2}\) from the state-operator map. This follows from the fact that any correlator involving a descendent field \(L_{-n}\mathcal{O}\) can be computed involving the corresponding primary field \(\mathcal{O}\) by applying a differential operator \(\mathcal{L}_{-n}\) in the following way:
\begin{align}
    &\braket{L_{-n}\mathcal{O}(z)....\mathcal{O}(z_i)}=\mathcal{L}_{-n}\braket{\mathcal{O}(z)....\mathcal{O}(z_i)} \nonumber \\ &\mathcal{L}_{-n}=\sum_{i=1}^{N}\left(\frac{(n-1)h_i}{(z_i-z)^n}-\frac{1}{(z_i-z)^{n-1}}\partial{z_i}\right)
\end{align}
This is a consequence of the stress-tensor Ward identities. These equations are called \textit{Belavin-Polyakov-Zamolodchikov equations} or \textit{BPZ} equations \cite{BELAVIN1984333}. One may obtain closed-form expression for a general \(rs^{th}\) order differential equation for correlators \(\mathcal{O}_{r,s}(z)\). The singular vector equation for a primary of conformal weight \(h_{r,1}\) will be,
\begin{equation}
    \sum_{\substack{p_i \geq 1 \\ p_1+...+p_k=r}} \frac{[(r-1)!]^2 (b^2)^{r-k}}{\prod_{i=1}^{k-1}(p_1+...+p_i)(r-p_1+...-p_i)}L_{-p_1}...L_{-p_k}\ket{h_{r,1}(b^2)} \label{Ldiff}
\end{equation}
where is the sum is over the partitions of \(r\) into \(k\) positive integers \(p_i\). \\ At large \(c\), the degenerate block dimension \(h_{r,s}\) becomes,
\begin{equation}
    h_{r,s}(c) \stackrel{c \rightarrow \infty}{\approx}\frac{c}{24}(1-r^2)+\frac{(1-s)}{2}+\frac{(1-r)(13+13r-12s)}{24}+\frac{3(r^2-s^2)}{2c}+....
\end{equation}
so \(h_{1,s}\) approaches a negative half-integer value at large \(c\), while \(h_{r,s}\) with \(r>1\) is \(\propto -c\). Alternatively, the operators with weight \(h_{1,s}\) can be understood of as light operator, while \(h_{r,s}\) corresponds to heavy operators and have a non-trivial effect on the geometry of AdS\(_3\) even in the semi-classical limit. These heavy operators lead to what is called \enquote{additional angles} in AdS\(_3\).
\section{Analytic Continuation and Unitarity}
\begin{figure}
    \centering
    \includegraphics[width=1.0\linewidth]{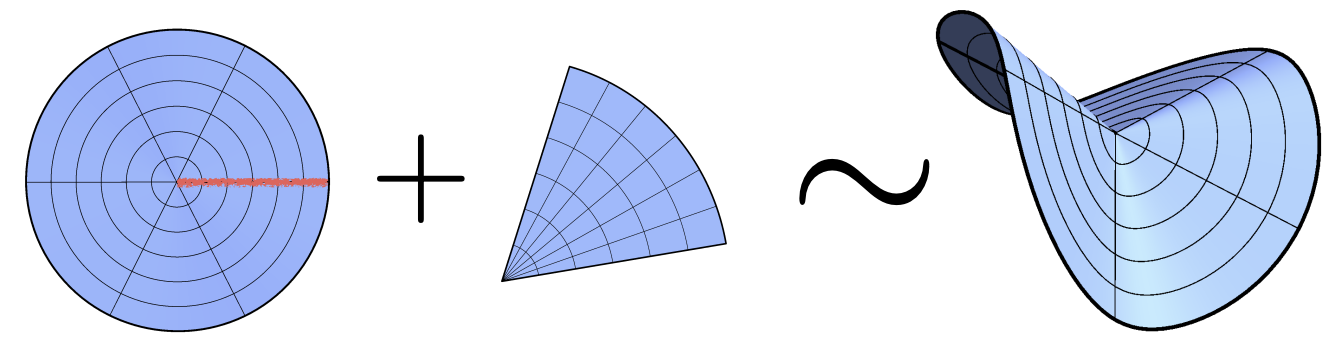}
    \caption{This depicts the surplus angle geometry generated from the insertion of a negative weight state in AdS\(_3\) \cite{Fitzpatrick_2017}.}
    \label{fig:surplus}
\end{figure}

We know that any state that is part of a unitary CFT must have a positive norm. In the case of CFT\(_2\), we require \(c > 0\) and \(h \geq 0\). This means that in the large \(c\) limit, states with conformal dimension \(h_{r,s}\) will not be unitary \footnote{We can have unitary CFT states with \(h_{r,s}>0\) in the limit \(c \rightarrow -\infty\) in dS/CFT \cite{Strominger_2001}.}. The consequence of this is that states with \(h_{r,1}\) act like large negative mass sources in AdS. Gravitational solutions incorporating such sources will have \(r_+=ir+O(\frac{1}{c})\) in the geometry of \eqref{metric}, implying that they have an angular surplus \(\Delta \phi\), for a total of \(2\pi r\) radians (depicted in Figure \ref{fig:surplus}). This contrasts with positive mass sources, which create deficit angles in the AdS background geometry \cite{DESER1984405, Fitzpatrick_2014, Fitzpatrick_2017}.
Let's consider a heavy operator \(\mathcal{O}_{2p+1,1}\). In the large \(c\) limit, the conformal weight of the operator will be,
\begin{equation}
    h_{2p+1,1}=\frac{c}{24}\left(1-(2p+1)^2\right)
\end{equation}
The surplus angle produced by a heavy operator \(\mathcal{O}_{2p+1,1}\) will be,
\begin{equation}
    \Delta \phi=2\pi\left(1-\sqrt{1-\frac{24h_{2p+1,1}}{c}}\right)=4\pi p
\end{equation}
We know that Virasoro blocks are meromorphic functions with only simple poles. The Zamolodchikov recursion relations in the \(z\) and \(q\)-series expansion of the blocks are based in this property \cite{Zamolodchikov1984,Alkalaev_2016}. It means that as a function of \(h_L\) and \(h_H\) are completely analytic. This is due to the fact that the \(q\)-expansion of the blocks converges away from OPE limit \cite{maldacena2015lookingbulkpoint}, and the coefficients in the \(q\)-series expansion are rational functions of \(c\) and polynomials in \(h_L\) and \(h_H\). In the large \(c\) limit, the degenerate conformal blocks have forbidden singularities, which are related to the forbidden singularities that arise from Euclidean time periodicity by an analytic continuation. This provides strong evidence in favour of the that the degenerate blocks \enquote{know} about the information restoring effects. 
\section{Forbidden Singularity: An Example}
In this section, we will provide a simple example to illustrate how forbidden singularities appear at large \(c\), and how their removal depends on non-perturbative effects. We will be studying an exact holomorphic Virasoro vacuum block \(\mathcal{V}_{2,1}\) which comes from a four-point function \(\braket{\mathcal{O}_{2,1}(\infty)\mathcal{O}_L(0)\mathcal{O}_L(z)\mathcal{O}_{2,1}(1)}\), where \(\mathcal{O}_{2,1}\) is the heavy operator in the large \(c\) limit since,
\begin{equation}
    h_{2,1}=-\frac{1}{4}(3b^2+2)
\end{equation}
goes to minus infinity as \(c \equiv b^2 \rightarrow \infty\). In this limit, the heavy operator produces a additional angle of \(2\pi\). This surplus angle geometry in AdS\(_3\) leads to forbidden singularities at \(z=2\) in the CFT vacuum block. For convenience, we take \(h_L=1\). The null condition for the primary \(\mathcal{O}_{2,1}\) is,
\begin{equation}
    \left[\frac{1}{b^2}\partial^2_z+\frac{1}{(1-z)^2}+\left(\frac{2}{b^2 z}+\frac{(2-z)}{z(1-z)}\right)\partial_z \right]\Tilde{\mathcal{V}}_{2,1}(z)=0 \label{diff}
\end{equation}
where we define \(\Tilde{\mathcal{V}}_{2,1}=z^{2h_{2,1}}\mathcal{V}_{2,1}\). The solution of this 2nd order differential equation corresponds to the vacuum block,
\begin{equation}
    \Tilde{\mathcal{V}}_{2,1}(b,z)=(1-z)\ _2F_1(2,b^2+1,2b^2+2,z)
\end{equation}
we see that \(\Tilde{\mathcal{V}}_{2,1}(b,2) \propto b^2\) and becomes singular in the large \(c \equiv b^2\) limit. To see the emergence of the forbidden singularities, we directly take \(b \rightarrow \infty\) limit in \eqref{diff}, after which we find,
\begin{equation}
    \left[1+\frac{(2-z)(1-z)}{z}\partial_z \right]\Tilde{\mathcal{V}}_{2,1}(\infty,z)=0 \label{diff2}
\end{equation}
whose solution is,
\begin{equation}
    \Tilde{\mathcal{V}}_{2,1}(\infty,z)=\frac{1-z}{(2-z)^2}
\end{equation}
As expected, we have a forbidden singularity at \(z=2\), and this singularity has the same nature as the \(z^{-2}\) OPE singularity of \(\mathcal{V}_{2,1}\). By comparing \eqref{diff} and \eqref{diff2}, we notice that the non-perturbative corrections in the vacuum block originate from the \enquote{perturbative} corrections to the differential equations that it obeys.
\section{Degenerate Heavy States} \label{DHS}
In this section, we will deal with degenerate states with conformal dimension \(h_{r,1}\). The vacuum block associated to such states will have forbidden singularities at \(c \rightarrow \infty\) and a non-trivial non-perturbative expansion in \(1/c\). \\ Suppose we have \(h=h_{r,1}\), then the BTZ horizon radius becomes \(r_+=2\pi r\) for any positive integer \(r\), the large \(c\) vacuum block defined on a plane becomes,
\begin{equation}
    \Tilde{\mathcal{V}}(t) \equiv \frac{e^{h_L t}(1-e^{-t})^{2h_L}}{(\sinh\left({\frac{rt}{2}}\right))^{2h_L}}
\end{equation}
where this function has singularities at \(t=\frac{\pi ik}{r}\) for \(k=0,1,...,r-1\) and the case when \(k=0\) corresponds to the OPE limit, whereas other singularities are forbidden. \\ The differential equation for this heavy-light vacuum block when \(h_H=h_{2,1}\) is obtained by from the large \(c\) limit of the \(r^{th}\) order differential equation obtained from the operator equation \eqref{Ldiff}. We note that in the large \(c\) limit, the states \(\mathcal{O}_{r,1}\) having conformal dimension \(h_{r,1}\) are annihilated by,
\begin{equation}
    0=\left(L_{-r}+\frac{1}{c}\sum_{p_i}b_{p_i}L_{-p_1}...L_{-p_k}\right)\ket{h_{r,1}}
\end{equation}
Making this differential equation act on a four-point function, we get,
\begin{equation}
    0=\braket{h_{r,1}|L_{r}\mathcal{O}_{r,1}(0)\mathcal{O}_{L}(x)\mathcal{O}_{L}(y)}+\sum_{p_i}\frac{b_{p_i}}{c}\braket{h_{r,1}|L_{p_k}...L_{p_1}\mathcal{O}_{r,1}(0)\mathcal{O}_{L}(x)\mathcal{O}_{L}(y)}
\end{equation}
Note that all the \(L's\) can be commuted to the right till they annihilate the vacuum. They all commute with \(\mathcal{O}_{r,1}(0)\), since it's a primary at the origin, while the commutators with \(\mathcal{O}_{L}\) produce \(h_L\) as their eigenvalue. As a consequence only \(L_r\) contributes at leading order in \(1/c\). The four-point function in our present case, which is related to \(\mathcal{V}(z)\) can be written in the following way,
\begin{equation}
    \braket{h_{r,1}|L_{r}\mathcal{O}_{r,1}(0)\mathcal{O}_{L}(x)\mathcal{O}_{L}(y)}=\frac{1}{(x-y)^{2h_L}}\Tilde{\mathcal{V}}\left(1-\frac{x}{y} \right) \label{Vz}
\end{equation}
Hence, the action of \(L_r\) on the four-point function takes the following form,
\begin{equation}
    0=x(x-y)(x^r-y^r)\Tilde{\mathcal{V}^{'}}\left(1-\frac{x}{y} \right)+yh_L \Tilde{\mathcal{V}}\left(1-\frac{x}{y} \right)\left(x^r(-rx+ry+x+y)-y^r(rx-ry+x+y)\right)
\end{equation}
By setting \(y=1\) and \(x=e^{-t}\), the above equation reduces to 
\begin{equation}
    \left(\partial_t -h_L g_r(t)\right)\Tilde{\mathcal{V}}(t)=0 \label{grt}
\end{equation}
where,
\begin{equation}
    g_{r}(t)=\coth \left({\frac{t}{2}} \right)-r\coth \left({\frac{rt}{2}} \right)
\end{equation}
The exact differential equation for a heavy-light vacuum block with operator with \(h_{2,1}\) can be written in an alternative form,
\begin{equation}
    \left(\partial_t -g_2 (t) \frac{h_L+b^{-2} \partial^2_t}{1+b^{-2}}\right)\Tilde{\mathcal{V}}(t)=0 
\end{equation}
which approaches \eqref{grt} in the limit \(b \rightarrow \infty\). It is worth noting that the relations we obtain from the external degenerate operators are exact, which enables us to study \enquote{heavy-heavy} correlators, where all external operators have conformal dimension \(h \propto c\) at large \(c\). However, at present we will constraint ourselves to heavy-light correlators in AdS\(_3\).
\section{Quasi-Normal Modes from Exact States in CFT}
In the heavy-light limit, for heavy operators whose mass lies above the BTZ, crossing symmetry implies that the heavy-light OPE must contains a dense spectrum of states. The spectral distribution function contains poles at the locations of the quasi-normal modes for a given BTZ metric \footnote{Such modes are unstable and contain imaginary components in the frequency, so they do not correspond to any primary state in a CFT, for which we must require stable eigenstates.}\cite{Birmingham_2002}. \\ From the Coulomb Gas formalism for degenerate , we know that for an anomalous OPE, the modified conformal dimension for the vertex operator is 
\begin{align}
    &h(\alpha)=\alpha(Q-\alpha) \label{halp} \nonumber \\ &Q = \left(b+\frac{1}{b}\right) 
\end{align}
where we define \(\alpha\)as the charge of the vertex operator  and \(Q\) as background charge, which makes \(U(1)\) symmetry anomalous. A physical operator of conformal dimension \(h_(\alpha)=\alpha Q-\alpha^2\) is associated to with the vertex operators \(V_{\alpha}\) and \(V_{Q-\alpha}\). The admissible charges for the degenerate vertex operators will be 
\begin{equation}
    \alpha_{r,s}=\frac{(1-r)b}{2}+\frac{(1-s)b^{-1}}{2} \label{alpha}
\end{equation}
and \(\alpha_{\pm}\) correspond to the charges screening operators \(\mathcal{O}_{\pm}\). These screening operators play an essential role in conserving the \textit{neutrality condition} in a vertex operator OPE. A keen observation of \eqref{alpha}, tells us \(\alpha_+ \equiv b \;\ \text{and} \;\ \alpha_- \equiv 1/b\). \\ Now we want to understand the fusion between degenerate operator \(\mathcal{O}_{r,s}\) and degenerate operator \(\mathcal{O}_{H}\), for which we make use the fusion rule of primaries \cite{DiFrancesco:1997nk},
\begin{equation}
        \mathcal{O}_{(r,s)} \times \mathcal{O}_{(\alpha)}= \sum_{\substack{k=1-r \\ k+r=1(\bmod{2})}}^{k=r-1} \;\ \sum_{\substack{l=1-s \\ l+s=1(\bmod{2})}}^{l=s-1} \mathcal{O}_{(\alpha+k\alpha_+ + l\alpha_-)} 
\end{equation}
for the given values of \(p,q\),
\begin{align}
    &q=-(s-1),-(s-3),...,(s-3),(s-1) \nonumber \\ &p= -(r-1),-(r-3),...,(r-3),(r-1)
\end{align}
This helps us write the fusion of degenerate light operator \(\mathcal{O}_{1,s}\) with an the operator \(\mathcal{O}_H\) having charge \(\alpha_H\), 
\begin{equation}
    \mathcal{O}_{(1,s)} \times \mathcal{O}_{(\alpha)}= \sum_{\substack{\frac{q}{2}=1-s \\ \frac{q}{2}+s=1(\bmod{2})}}^{\frac{q}{2}=s-1} \mathcal{O}_{(\alpha+\frac{q}{2}b^{-1})}
\end{equation}
and the fused states can only have charge given by,
\begin{align}
        &\alpha_b=\alpha_H+\frac{qb^{-1}}{2} \nonumber \\ \forall \;\ &q=-(s-1),-(s-3),...,(s-3),(s-1) 
\end{align}
To obtain an expression for \(\alpha_b\), we make use of the following identities,
\begin{align}
    &h_H=\frac{c}{24} \left(1-(2\pi iT_H)^2 \right) \nonumber \\ &Q=\left(b+\frac{1}{b} \right) = \sqrt{\frac{c-1}{6}}
\end{align}
Now working in the large \(c\) limit, we can expand \(Q\) as 
\begin{equation}
    Q=\sqrt{\frac{c}{6}}\left(1-\frac{1}{2c}+....\right) \label{Qc}
\end{equation}
Solving the quadratic equation and replacing \(b^2=c/6\) with \(h_H/c\) fixed, we get
\begin{equation}
    \alpha_H \approx \frac{b}{2}(1 \pm 2\pi iT_H)+\frac{2}{b}(1 \mp \frac{1}{24\pi iT_H}+...)
\end{equation}
The spectrum of operators from in the fusion \(\mathcal{O}_{1,s} \times \mathcal{O}_{H}\) at large \(c\) is \cite{Turiaci_2016,fitzpatrick2016informationlossads3cft2},
\begin{equation}
    h_b=h_H+2\pi iT_H \left(h_{1,s}+n\right) \;\ , \;\ n=0,...,(r-1)
\end{equation}
This is the quasi-normal spectrum of the BTZ black hole, truncated at \(n=(r-1)\) for any integer \(r\).
\section{Resolving the Forbidden Singularities} \label{Forbidden}
We have established that at large \(c\), the heavy-light conformal blocks have forbidden singularities which persist in all orders of the \(1/c\) expansion. Since for any finite value of \(c\) the vacuum block must only have OPE singularities, we conclude that the forbidden singularities must be resolved by non-perturbative \(e^{-c}\) effects. In this section, we will review how these forbidden singularities are resolved at finite \(c\) by the exact Virasoro blocks. In order to understand how forbidden singularities are resolved in the degenerate state differential equation, we will work upto the sub-leading order in the large \(c\) equation \eqref{grt}. The null descendent equation upto the sub-leading order in \(1/c\) expansion is,
\begin{equation}
    0=\ket{\psi}=\left(L_{-r}+\frac{6}{c}\sum_{j=1}^{r-1}\frac{1}{j(r-j)}L_{-r+j}L_{-j}\right)\ket{h_{r,1}}
\end{equation}
As before, we let this differential operator act on the four-point correlation function,
\begin{align}
   &0= \braket {\psi|\mathcal{O}_{r,1}(0)\mathcal{O}_{L}(x_1)\mathcal{O}_{L}(x_2)} \nonumber \\ &\approx \braket{h_{r,1}|\mathcal{O}_{r,1}\left(L_{-r}+\frac{6}{c}\sum_{j=1}^{r-1}\frac{1}{j(r-j)}L_{-r+j}L_{-j} \right)\mathcal{O}_{L}(x_1)\mathcal{O}_{L}(x_2)}
\end{align}
and the four-point function is related to \(\mathcal{V}(z)\) via \eqref{Vz}. We are interested in the behaviour of the correlator near the forbidden singularity \(z=1-e^{\frac{2\pi in}{r}}\). To explore the behaviour near this singularity, we take \(x_1=1\) and \(x_2=1-z\) in a scaling limit,
\begin{equation}
    1-z=e^{-\frac{2\pi in}{r}-\frac{x}{b}}
\end{equation}
where \(b \rightarrow \infty\). At fixed \(|x|\) and large \(b\), this scaling limit allows us to explicitly see how singularities are removed by the finite \(c\) effects. Under this scaling limit, the differential equation we get is of the form,
\begin{equation}
    0=2h_L \mathcal{V}(x)+x\mathcal{V}^{'}(x)-\mathcal{V}^{''}(x)\sigma^2_n (r) \label{2diff}
\end{equation}
where 
\begin{equation}
    \sigma^2_n (r) \equiv 4\sum_{j=1}^{r-1}\frac{\sin^2\left({\frac{jn\pi}{r}}\right)}{rj(r-j)}
\end{equation}
At large \(r\), \(\sigma^2_n (r)\) takes a simple form,
\begin{equation}
    \sigma^2_n (r) \stackrel{r \gg 1}{\approx} \frac{4}{r^2}\int_0^{2\pi n} \frac{dt}{t}\sin^2{\left(\frac{t}{2}\right)}
\end{equation}
indicating that 
\begin{equation}
     \sigma^2_n (h_H) {\approx} -\frac{c}{6h_H}\int_0^{2\pi n} \frac{dt}{t}\sin^2{\left(\frac{t}{2}\right)}
\end{equation}
The solution of this differential equation is the confluent hypergeometric function \(_1F_1(h_L,\frac{1}{2}; \frac{x^2}{2\sigma^2_n(r)})\). We conjecture that this integral function solves the 2nd order differential equation \eqref{2diff},
\begin{equation}
    S(x,c) \approx \int_0^{\infty}dp p^{2h_L-1}e^{-px-\frac{\sigma^2}{2b^2}p^2} \label{S}
\end{equation}
Thus the function \(\sigma^2_n (r)\) sets the width of the correlator of order \(\frac{1}{\sqrt{h_{L}c}}\) in the \(z\) or \(t\) coordinates. We conjecture that this function characterises the general Virasoro vacuum block in the vicinity of the forbidden singularities at large but finite \(c\). \\ Let's provide some evidence on the choice of the integral solution \eqref{S} for the Virasoro vacuum block. The \(1/c\) expansion of this solution takes the form 
\begin{equation}
    \frac{1}{x^{2h_L}}-\frac{6\sigma^2 h_L(h_L+1)}{c}\frac{1}{x^{2h_L+2}}+....
\end{equation}
This accurately predicts the strength of the OPE singularity of \(x\) and the relation between \(h_L\) and \(h^2_L\). We have the expression for the heavy-light vacuum block upto the first sub-leading level \cite{Beccaria_2016,fitzpatrick2016conformalblockssemiclassicallimit}, we can extract the exact form of \(\sigma^2\) from the coefficient of \(x^{-2h_L-2}\). \\ The Virasoro vacuum block with the \(1/c\) corrections are \cite{fitzpatrick2016conformalblockssemiclassicallimit, fitzpatrick2016informationlossads3cft2},
\begin{align}
   \mathcal{V}(t)= &e^{h_L t}\left(\frac{\pi T_H}{\sin{\pi T_H t}}\right)^{2h_L}\left[1+\frac{h_L}{c}\mathcal{V}^{(1)}_{h_L}(t)+\frac{h^2_L}{c}\mathcal{V}^{(1)}_{h^2_L}(t)\right], 
   \nonumber \\ \mathcal{V}^{(1)}_{h_L}(t)= &\frac{\csch^2{(\frac{\alpha t}{2}})}{2}\Bigl[3(e^{-\alpha t}B(e^{-t},-\alpha,0)+e^{\alpha t}B(e^{-t},\alpha,0)+e^{-\alpha t}B(e^{t},\alpha,0)+e^{\alpha t}B(e^{t},-\alpha,0) \nonumber \\ &+\frac{1}{\alpha^2}+\cosh{(\alpha t)}\left(-\frac{1}{\alpha^2}+6H_{-\alpha}+ 6H_{\alpha}+6i \pi -5\right)+12\log\left(2\sinh\left({\frac{t}{2}}\right)\right)+5 \Bigr] \nonumber \\ &-t\frac{(13\alpha^2 -1)\coth\left({\frac{\alpha t}{2}}\right)}{\alpha}+12\log \left(\frac{2\sinh\left({\frac{\alpha t}{2}}\right)}{\alpha}\right) \nonumber \\ \mathcal{V}^{(1)}_{h^2_L}(t)= &\Bigl(6\csch^2 \left({\frac{\alpha t}{2}} \right) \Bigl[\frac{B(e^{-t},-\alpha,0)+B(e^{-t},\alpha,0)+B(e^{t},-\alpha,0)+B(e^{t},\alpha,0)}{2} 
 \nonumber \\ &+H_{-\alpha}+H_{\alpha}+2\log\left(2\sinh \left({\frac{t}{2}}\right)\right)+i\pi \Bigr] +2\left(\log\left(\alpha \sinh \left({\frac{t}{2}}\right)\csch^2\left({\frac{\alpha t}{2}}\right)\right)\right)+1\Bigr) \label{Vsl}
\end{align}
where \(B(x,\beta,0)=\frac{x^{\beta}\ _2F_1(1,\beta,1+\beta;x)}{\beta}\) is the incomplete Beta function, \(z=1-e^{-t}\), \(H_n\) is the Harmonic function and \(\alpha=\sqrt{1-\frac{24h_H}{c}} = 2\pi iT_H\). If we naively expand any of these functions around \(t=0\), they are non-singular. However, after the analytic continuation \(t \rightarrow t+\frac{n}{T_H}\), the singularities develop. We consider the forbidden singularities of the sub-leading terms of the heavy-light Virasoro block. So we evaluate, \eqref{Vsl} around \(t=\frac{n}{T_H}=\frac{2\pi in}{\alpha}\), which gives a coefficient for the \(n^{th}\) forbidden singularity \(1/(t-\frac{n}{T_H})^{2h_L+2}\),
\begin{align}
    \mathcal{V}^{(1)}_{h^2_L}= &2 \mathcal{V}^{(1)}_{h_L} \rightarrow -\frac{3}{2\pi^2 T^2_H}\Bigl[2H_{-\alpha}+2H_{\alpha}+2\pi i+4\log\left(2\sinh \left({\frac{n}{2T_H}}\right)\right) \nonumber \\&B(e^{\frac{n}{T_H}},\alpha,0)+B(e^{-\frac{n}{T_H}},\alpha,0)+B(e^{\frac{n}{T_H}},-\alpha,0)+B(e^{-\frac{n}{T_H}},-\alpha,0)\Bigr]
\end{align}
The degenerate blocks have forbidden singularities characterised by the function,
\begin{equation}
    \sigma^2_{deg}(n,r)=\frac{12\left(B(e^{-\frac{2\pi in}{r}},r,0)+B(e^{\frac{2\pi in}{r}},r,0)+2\log\left(2\sinh \left({\frac{\pi in}{r}}\right)\right)+2H_{r-1} \right)}{r^2}
\end{equation}
Using the identity of the Harmonic function \(H_{r-1}=H_r -\frac{1}{r}\), and analytically continuing \(r \rightarrow \alpha=2\pi iT_H\) we see that,
\begin{align}
    \frac{\sigma^2_{deg}(n,\alpha)+\sigma^2_{deg}(n,-\alpha)}{2}= &-\frac{3}{2\pi^2 T^2_H}\Bigl[2H_{-\alpha}+2H_{\alpha}+2\pi i+4\log\left(2\sinh \left({\frac{n}{2T_H}}\right)\right) \nonumber \\&+B(e^{\frac{n}{T_H}},\alpha,0)+B(e^{-\frac{n}{T_H}},\alpha,0)+B(e^{\frac{n}{T_H}},-\alpha,0)+B(e^{-\frac{n}{T_H}},-\alpha,0)\Bigr]
\end{align}
We conclude that the exact Virasoro blocks match the results of the \(1/c\) corrections to the heavy-light conformal block.
\section{Late Lorentzian Time Behaviour of Exact Blocks}
The AdS correlators in a black hole geometry background decay exponentially at late Lorentzian times, which signals information loss. We have shown that to leading order in large \(c\), heavy degenerate Virasoro blocks obey a 1st order differential equation in Section \ref{DHS}. However, there exists a universal 2nd order differential equation whose solutions incorporates both perturbative and non-perturbative effects in the \(1/c\) series expansion,
\begin{equation}
    h_L g_r(t)\frac{\mathcal{V}(t)}{\mathcal{V}^{'}(t)}-1=\frac{6}{c}\Sigma_H (t)\frac{\mathcal{V}^{''}(t)}{\mathcal{V}^{'}(t)} \label{uni}
\end{equation}
where \(t=t_E +it_L\) is the Euclidean time and the functions are,
\begin{align}
    &g_{r}(t)=\coth \left({\frac{t}{2}} \right)-r\coth \left({\frac{rt}{2}} \right) \nonumber \\ &\Sigma_r (t)=-\frac{1}{\sinh{\left(\frac{rt}{2}\right)}}\left(e^{-\frac{rt}{2}}\Tilde{B}_r (t)+e^{\frac{rt}{2}}\Tilde{B}_{r} (-t)-2\cosh{\left(\frac{rt}{2}\right)}\Tilde{B}_{r} (0) \right) \nonumber \\ &\Sigma_H (t)=\Sigma_r (t) +\Sigma_{-r}(t)
\end{align}
There are many different ways to represent the function \(\Tilde{B}_r (t)\),
\begin{align}
    &\Tilde{B}_r (t)=\sum_{j=1}^{r-1}\frac{e^{tj}}{j} \nonumber \\ &\Tilde{B}_r (t)=-\log(1-e^t)-\frac{e^{rt}\ _2F_1(1,r,1+r;e^{t})}{r}=-\log(1-e^t)-B(e^t;r,0) \nonumber \\ &e^{-t}\frac{d}{dt}\Tilde{B}_r (t)=\frac{1-e^{t(r-1)}}{1-e^t}
\end{align}
where \(B(x;a,b)\) is the incomplete Beta function. We can use \eqref{uni} to analytically continue \(r \rightarrow 2\pi iT_H\) \footnote{We identify the parameter \(r=2\pi iT_H=\sqrt{1-\frac{24h_H}{c}}\) where \(T_H\) is the Hawking temperature associated with the heavy operator.} to study correlation functions associated to BTZ black hole. Increasing \(\mathrm{Im}\ t\) is equivalent to a rotation on a unit circle, i.e the function \(e^t\) winds around 1 in the complex \(t\)-plane, and picking up a contribution each time from the pole at \(t=0\). This allows the function to keep track of how much Lorentzian time has passed. 
\begin{figure}[t]
    \centering
    \includegraphics[width=0.95\linewidth]{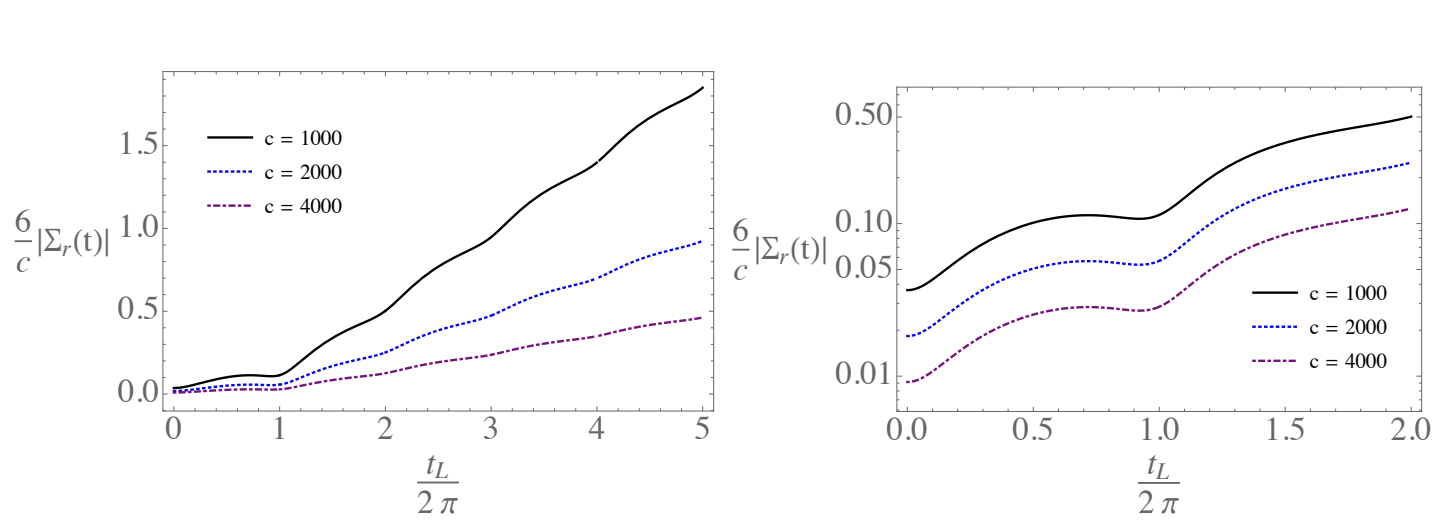}
    \caption{We take \(r=2\pi iT_H=\frac{i}{2}\) , and time \(t=t_E+it_L\) with constant \(t_E=-i\). \textit{Left}: the plot shows the magnitude of the coefficient term \(\frac{6}{c}\Sigma_r (t)\) for various values of \(c\) as a function of the Lorentzian time. At late Lorentzian times, the slope is \(\sim O(1/S_{BH})\). \textit{Right}: the left plot drwan on the \(\log\) scale \cite{fitzpatrick2016informationlossads3cft2}.}
    \label{fig:enter-label}
\end{figure}
In the semi-classical large \(c \propto b^2\), we can ignore \(\mathcal{V}^{''}\) term to obtain the differential equation for the heavy-light Virasoro vacuum block. However, when \(\mathcal{V} \sim O(c)\) in the vicinity of the forbidden singularity, non-perturbative effects come into play. We can obtain \eqref{2diff} can be obtained by the universal differential equation by scaling towards a forbidden singularity and taking the large \(b\) limit with \(x=bz\) fixed. \\ As we increase the Lorentzian time \(t_L=\mathrm{Im}\ t\), it shifts the value \(\Tilde{B}_r (t)\) by an exponent every \(2\pi\). This produces a linear growth in \(\Sigma_r\) for late Lorentzian times. Due to the linear growth of \(\Sigma_r (t)\) as a function of \(t_L\) and the parametrisation \(r=2\pi iT_H\) we find,
\begin{equation}
    \frac{1}{c}\Sigma_r (t) \sim \left(\frac{t_L}{cT_H}\right) \sim \left(\frac{t_L}{S_{BH}}\right)
\end{equation}
where the black hole entropy \(S_{BH}=\frac{\pi^2}{3}cT_H\). This proves that at late Lorentzian times \((t_L \sim S_{BH}) \) there is a change of behaviour from the exponential decay of the Virasoro blocks 

\chapter{Information Restoration}
In this section, we will explicitly show that information loss is resolved by
non-perturbative effects. We will take two approaches, one using the Borel resummation due to its close connection to the classical solutions of the field equations. While in the alternative approach, we will describe exact 
conformal blocks as the Coulomb Gas contour integrals. Our motivation is to connect the behavior of the Virasoro vacuum block to the saddle points and the contour of integration for the Chern-Simons path integrals in AdS\(_3\) \cite{witten2010analyticcontinuationchernsimonstheory,Gaiotto_2012}.
\section{Borel Resummation} \label{borel}
We wish to understand the resolution of information loss from a perturbative series in \(G_N=\frac{3}{2c}\), i.e. we wish to expand the Virasoro conformal block as,
\begin{equation}
    \mathcal{V}(z)=\mathcal{V}_{c=\infty}(z)\left(1+\frac{f_1(z)}{c}+...\right)+e^{-cs(z)}\left(g_0(z)+\frac{g_1(z)}{c}+...\right)+...
\end{equation}
The first term in the perturbative series is the AdS\(_3\) vacuum block, while the other terms denote the non-perturbative effects. \\ Suppose we have a formal power series,
\begin{equation}
    f(g)=\sum_n a_{n}g^n
\end{equation}
we can define a Borel sum \(\mathcal{B}(g)\) by \(a_n \rightarrow  \frac{a_n}{n!}\), where the Borel transformation is its equivalent exponential series. Then the Borel sum is defined as,
\begin{equation}
    f(g)=\int_0^{\infty}e^{-t}\mathcal{B}(tg)dt=\int_0^{\infty}\frac{dy}{g}e^{-\frac{y}{g}}\mathcal{B}(y)
\end{equation}
If the Borel integral converges and has no singularities on the real axis, then it can be viewed as the definition of \(f(g)\). Singularities on the Borel plane lead to branch cuts when we analytically continue \(f(g)\) \cite{Basar_2013}. In his paper \cite{tHooft1979} 't Hooft argued that singularities on the Borel plane can be viewed as the classical solutions of the field equations by equating the Borel integral and the path integral form of the correlator,
\begin{equation}
    \int_0^{\infty}dy \ e^{-\frac{y}{g}}\mathcal{B}(y) \sim \int \mathcal{D}\phi \ e^{-\frac{1}{g}S(\phi)}
\end{equation}
which leads to,
\begin{align}
    \mathcal{B}(y) \sim &\int \mathcal{D}\phi \ \delta(y-S(\phi)) \nonumber \\ &\sim \left(\frac{dS}{d\phi}\right)^{-1} \Bigg|_{S(\phi)=y}
\end{align}
Thus the Borel sum has a singularity at some \(y_*\) such that,
\begin{equation}
    \frac{dS}{d\phi_*}=0 \;\ , \;\ S(\phi_*)=y_*
\end{equation}
The singularities of \(\mathcal{B}(y)\) in the \(y\) plane correspond to solutions of the classical equations of Einstein's equations in AdS\(_3\) with an action \(y_*\). Now have established the premise for the Borrel sum, we go to back to resolving the singularities. Forbidden singularities take the form of power laws \(x^{-2h}\) and we want a model that regulates the singularities in a way that in the limit \(c \rightarrow \infty\), we recover the power law form but at finite \(c\), we have the entire function of \(x\). Let's consider the Borel resummation of our universal function \eqref{S} as a \(1/c\) series expansion,
\begin{align}
    S(x,c)=&\int_0^{\infty}dp \ p^{2h_L-1} e^{-px}\sum_{n=0}^{\infty}\frac{1}{n!}\left(-\frac{1}{c}\right)^{n}p^{2n} \nonumber \\ &\sum_{n=0}^{\infty}\frac{\Gamma(2h+2n)}{n!}\left(-\frac{1}{c}\right)^{n} x^{-2h-2n}
\end{align}
We define a Borel transform of the \(1/c\) perturbation series,
\begin{align}
    \mathcal{B}(x,y)=&\sum_{n=0}^{\infty}\frac{\Gamma(2h+2n)}{n!^2}(-y)^{n} x^{-2h-2n} \nonumber \\ =&\frac{\Gamma(2h)}{x^{2h}}\ _2F_1(h,h+\frac{1}{2},1,-\frac{4y}{x^2})
\end{align}
and we expect that the original function \(S(x,c)\) can be written as,
\begin{equation}
    S(x,c)=\int_0^{\infty} dy \ e^{-y}\mathcal{B}(x,\frac{y}{c})
\end{equation}
This is only true when there are no singularities on the real \(y\)-axis. However, the hypergeometric function \(_2F_1\) has branch cuts extending from 1 to \(\infty\). From this we can infer that the Borel integral has a branch cut starting at,
\begin{equation}
    y=-\frac{x^2 c}{4}
\end{equation}
and extending to \(\infty\). This will intersect the real axis when e.g. \(c>0\) and real, while \(x\) is imaginary. In the context of the Virasoro blocks, we would take \(x=z-z_*\), with \(z_*\) being the location of the forbidden singularity, making the imaginary values of \(x\) important. Furthermore, when \(x \approx 0\) with fixed \(c\), we are in the vicinity of the forbidden singularity. In this region, the Borel resummation breaks down, signaling the importance of non-perturbative effects near such singularities.
\section{Heavy-light Vacuum Block}
A heavy-light vacuum block with heavy degenerate states \(h_{2,1}\) and light states \(h_L\) is defined by,
\begin{equation}
    \Tilde{\mathcal{V}}(z)=z^{2h_{2,1}}\mathcal{V}(z)=(1-z)\ _2F_1(2,b^2+1,2b^2+2;z)
\end{equation}
Our goal will be simple; study the \(1/b^2\) series expansion of the vacuum block which is a function of the kinematic variable \(t=-\log(1-z)\) and then Borel resum the resulting asymptotic series. The 2nd order differential equation that the block satisfied provides a recursion relation, which takes a simple form when \(h_L=1\). We define a new variable,
\footnote{We use the quadratic transformation property of the hypergeometric function \(_2F_1(a,b,2b;z)=(1-z)^{-\frac{a}{2}}\ _2F_1(\frac{a}{2},b-\frac{a}{2},b+\frac{1}{2};\frac{z^2}{4(z-1)})\), where we have defined \(e^{s}=-\frac{z^2}{4(z-1)}\)}
\begin{equation}
    s =2\log \left(\sinh\left({\frac{t}{2}}\right)\right)
\end{equation}
where the forbidden singularity at \(z=2\) corresponds to \(s=\pi i\). We can write the degenerate block as,
\begin{equation}
    \Tilde{\mathcal{V}}(s)=\sqrt{1+e^{-s}}\sum_{k}\frac{1}{b^{2k}}G_k (s)
\end{equation}
The coefficients can be easily found \footnote{We have used the identity 15.2.5 of \cite{10.5555/1098650}},
\begin{equation}
    G_k(s)=\left(-\partial_s+\frac{1}{2}\right)\left(-\partial_s\right)^{k-1}G_0(s)
\end{equation}
where the leading coefficient is,
\begin{equation}
    G_0(s)=\frac{e^{\frac{s}{2}}}{(1+e^s)^{\frac{3}{2}}}
\end{equation}
The Borel transform of the power series in the degenerate conformal block then becomes,
\begin{equation}
    \mathcal{B}(s,y)=\sum_{k}^{\infty}\frac{y^k}{k!}\left(-\partial_s+\frac{1}{2}\right)\left(-\partial_s\right)^{k-1}G_0(s)=-\frac{e^{\frac{3}{2}(s-y)}}{(1+e^{s-y})^{\frac{3}{2}}}+\frac{1}{(1+e^{-s})^{\frac{1}{2}}}
\end{equation}
This allows us to write the degenerate heavy-light block as a Borel transform,
\begin{equation}
    \Tilde{\mathcal{V}}_{2,1}(b,s)=1-\sqrt{1+e^{-s}}\int_0^{\infty}\frac{e^{-y-\frac{3y}{2b^2}+\frac{3s}{2}}}{(1+e^{-s+\frac{y}{b^2}})^{\frac{3}{2}}} \label{BRV}
\end{equation}
The integrand \eqref{BRV} has poles at,
\begin{equation}
    y_n=b^2 (s-\pi i(1+2n))
\end{equation}
for any integer \(n\). We know that when \(s \approx \pi i\), the Borel integral has a singularity at \(y=0\). This indicates that perturbative \(1/c\) series breaks down, which is what we expect in the vicinity of the forbidden singularity. Expanding \(s(t)\) around the forbidden singularity \(t=\pi i\), we get
\begin{equation}
    s(t) \approx \pi i+\frac{1}{4}(t-\pi i)^2+..
\end{equation}
To keep the singularities \(y_n\) fixed in the Borel plane as we take the semi-classical limit \(b \rightarrow \infty\), we need to keep \(b(t-\pi i)\) constant. This is the same scaling limit in the Section \ref{Forbidden}. As we explained before that the singularities in the Borel plane correspond to the solutions of Einstein's equations in AdS. We expect that the classical solutions \enquote{solitons} corresponds to the heavy, non-perturbative states of the theory. The degenerate Virasoro block \(\mathcal{V}_{2,1}\) represents a correlator of exact CFT operators \(\mathcal{O}_{2,1}\), the physical states produced from the fusion are,
\begin{equation}
    \mathcal{O}_{2,1} \times \mathcal{O}_{2,1}=\mathbb{1}+\mathcal{O}_{3,1}
\end{equation}
where the non-perturbative effects come from the heavy degenerate state \(\mathcal{O}_{3,1}\) while \(\mathbb{1}\) corresponds to the vacuum block. In the following sections, we will analyse some general examples such as \(\mathcal{O}_{3,1} \times \mathcal{O}_{3,1}\) and \(\mathcal{O}_{2,2} \times \mathcal{O}_{2,2}\) to better understand which heavy, exact states in the CFT correspond to the non-perturbative effects.
\section{Coulomb Gas Formalism for Exact Blocks}
The degenerate blocks can be written as contour integrals, which is formally known as the Coulomb gas formalism \cite{DiFrancesco:1997nk,DOTSENKO1984312,DOTSENKO1985691}. This means that we can study Virasoro blocks at large \(c\) using asymptotic analysis. The Coulomb Gas integral takes the following form,
\begin{equation}
    \mathcal{V}=\int_{C}dw \ e^{\mathcal{I}(b,z;w)}
\end{equation}
for some contour, we study the saddle points \cite{witten2010analyticcontinuationchernsimonstheory} of the action \(\mathcal{I}\) at large but finite \(c \propto b^2\). The Coulomb Gas formalism is connected to the semi-classical gravitational path integral in AdS\(_3\). There is ongoing research work to establish a relationship between the saddle points of the Coulomb gas integrals and classical solutions of AdS\(_3\) gravity, Chern-Simons theory, or Liouville theory. We will shortly see that the physical exchanged states in the conformal blocks do not correspond to a single saddle point in the \(\mathcal{I}\) but as a linear combination of saddle points. 
\subsection{Heavy-light Blocks with \(\mathcal{O}_{2,1}\)}
In this section we will consider a heavy-light degenerate Virasoro block with external heavy operators \(\mathcal{O}_{2,1}\) and light operators with \(h_L=1\). Then the four-point correlation function becomes,
\begin{equation}
    \braket{\mathcal{O}_{2,1}(\infty)\mathcal{O}_L(0)\mathcal{O}_L(z)\mathcal{O}_{2,1}(1)} \sim z^{-2h_{2,1}}(1-z)\int_{C}dw \ e^{\mathcal{I}} \label{cogas}
\end{equation}
and the action is,
\begin{equation}
    \mathcal{I}(w) \equiv b^2\log[w(1-w)]-2\log(1-wz) \label{I}
\end{equation}
The integral has singularities at \(w=\frac{1}{z}\). These singularities play a crucial role in the ensuring the monodromy invariance of the integral around the singular points. There are two natural choices of contours \([0,1]\) and \([\frac{1}{z},\infty)\) \footnote{Such contour integrals are also known as Pochhammer integrals \cite{DiFrancesco:1997nk}} on which the integral can be performed. \\ To connect a specific contour choice to the physical content of the conformal block, we can take the OPE limit \(z \rightarrow 0\). In the small \(z\), for the contour choice \([0,1]\), we get,
\begin{equation}
    z^{-2h_{2,1}}(1-z)\int_{0}^{1} dw \ e^{\mathcal{I}_{2,1}} \propto z^{-2h_{2,1}}
\end{equation}
which means that the this contour choice gives us the vacuum block. While the contour \([\frac{1}{z},\infty)\) produces,
\begin{equation}
    z^{-2h_{2,1}}(1-z)\int_{\infty}^{\frac{1}{z}} dw \ e^{\mathcal{I}_{2,1}} \propto  z^{-2h_{2,1}} \times z^{-2b^2-1}=z^{-2h_{2,1}+2h_{3,1}}
\end{equation}
\begin{figure}
    \centering
    \includegraphics[width=0.80\linewidth]{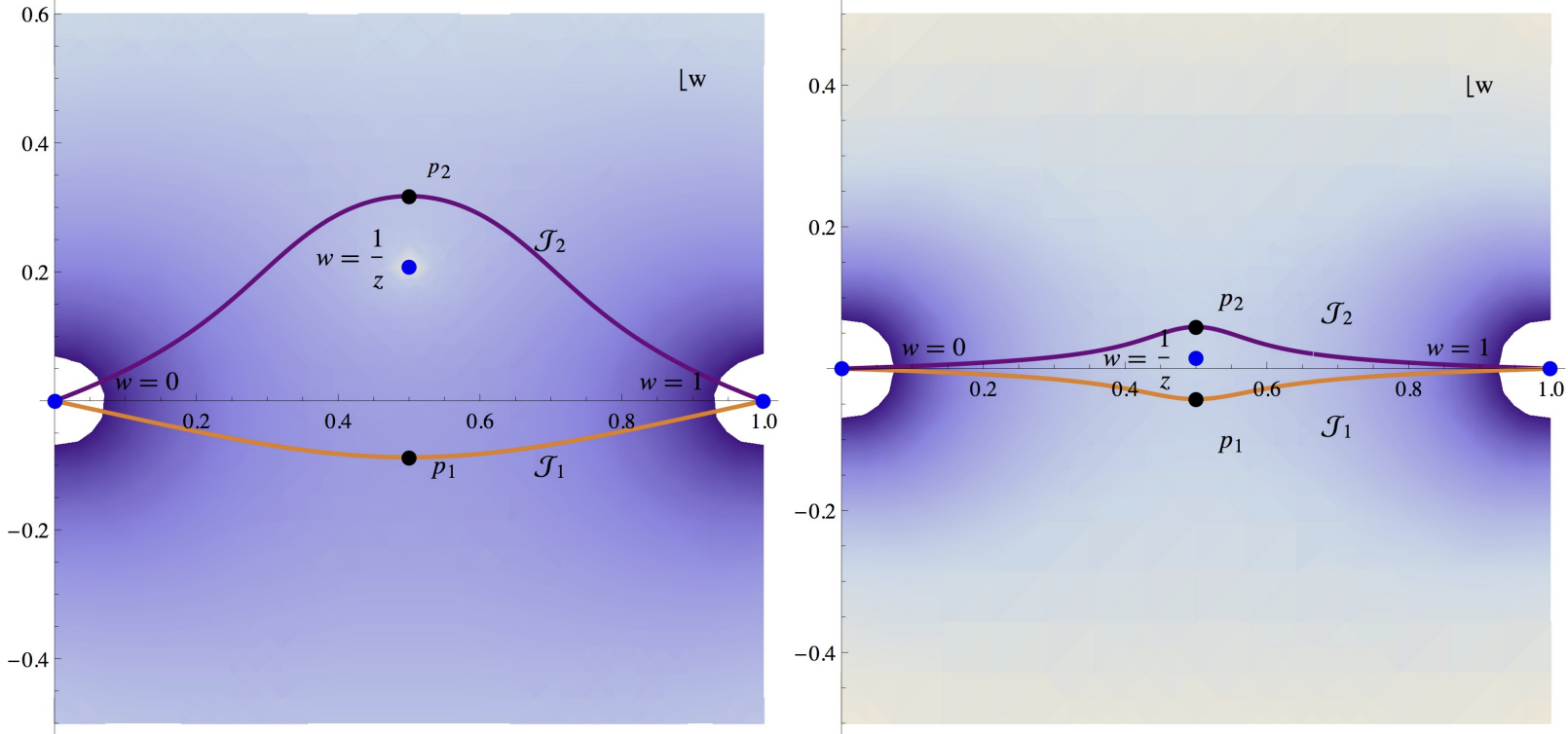}
    \caption{This figure displays the positions of the critical points (black dots) of the action \(\mathcal{I}_{2,1}\) along with it's respective steepest descent contour, with the  end points \(w=0,1\). This integrand has a singularity at \(w=\frac{1}{z}\), thus the two paths \(\mathcal{J}_1\) and \(\mathcal{J}_2\) cannot be deformed into each other. In the left figure, we have chosen \(b^2=10\) and \(z=1-e^{\frac{3\pi i}{4}}\), whereas in the right figure we have \(b^2=100\) and \(z=1-e^{0.98 \pi i}\). Via this we showcase that the two contours will approach each other as we go near the forbidden singularity which is at the point \(z=2\) \cite{fitzpatrick2016informationlossads3cft2}. }
    \label{fig:steep}
\end{figure}
This Virasoro block corresponds to the exchange of the heavy state \(\mathcal{O}_{3,1}\) and it's primary descendants. \\ We know that critical/saddle points occur when the action \(\mathcal{I}(w)\) is stationary with respect to the parameter \(w\). Each such critical point \(w=p_i\) is connected to steepest descent contours (the real part of the action \(\mathcal{I}(w)\)) passing through it. Following the steps of Ed Witten \cite{witten2010analyticcontinuationchernsimonstheory, Koshkarov1995}, we will refer the union of steepest descent contours passing through a critical point \(p_i\) as \(\mathcal{J}_i\), also known as \enquote{\textit{Lefschetz thimbles}}, which we will discuss once gain in Appendix \ref{ltsdc}. The steepest descent contour are curves in a complex \(w\) plane, parameterised by a real number \(t\) so that \(w(t)\) satisfies the flow equations,
\begin{equation}
    \frac{dw}{dt}=-\frac{\partial\bar{\mathcal{I}}(w)}{\partial \bar{w}} \;\ , \;\ \frac{d \bar{w}}{dt}=-\frac{\partial\mathcal{I}(w)}{\partial w} \label{flow}
\end{equation}
The \(\mathrm{Im} \mathcal{I}(w)\) is constant along the steepest contour,
\begin{equation}
    \frac{1}{2i}\frac{d(\mathcal{I}-\bar{\mathcal{I}})}{dt}=\frac{1}{2i}\left(\frac{\partial\mathcal{I}(w)}{\partial w} \frac{dw}{dt}-\frac{\partial\bar{\mathcal{I}}(w)}{\partial \bar{w}}\frac{d \bar{w}}{dt} \right)=0
\end{equation}
This directly follows from the flow equations \eqref{flow} and the chain rule. So the steepest descent contours can be evaluated using the condition \(\mathrm{Im} \ \mathcal{I}(w)=\text{constant}\). For our case, \(\mathcal{I}_{2,1}(w)\) has a singularity at \(w=\frac{1}{z},\infty\) and vanishes at \(w=0,1\), (i.e when \(\mathcal{I}\rightarrow -\infty\) the steepest contours must end at \(w=0,1\)). 
The action \(\mathcal{I}_{2,1}(w)\), \eqref{I} has two saddle points at \(w=p_1\) and \(p_2\). Since the forbidden singularities of a exact Virasoro block lie on a unit circle, we will write \(z=1-e^{i\theta}\) such that \(\frac{1}{z}=\frac{1}{2}+\frac{i}{2}\cot{\frac{\theta}{2}}\) and we will work in the real and large \(b^2 \gg 1\). In this parametrisation \(z\) , we can see that the pole of the integrand \eqref{cogas} cuts in the \(w\) contour \([0,1]\) each time \(\theta \rightarrow \theta+2\pi\), and this induces a monodromy in Virasoro vacuum block. The two critical points can be denoted by,
\begin{align}
    &p_1=\frac{1}{2}+iq_- \;\ , \;\ p_2=\frac{1}{2}+iq_+ \nonumber \\ &q_{\mp}=\frac{2b^2\cos\left({\frac{\theta}{2}}\right) \mp \sqrt{2(b^2-2)^{2}\cos{\theta}+2b^4+8b^2-8}}{8(b^2-1)\sin \left({\frac{\theta}{2}}\right)}
\end{align}
The critical points and the associated steepest contours are displaced in Figure \ref{fig:steep}. The function are related by \(q_+(+\pi)=q_-(-\pi)\), implying that the two critical points coincide at the forbidden singularity in the semi-classical limit \(b \rightarrow \infty\), which lies on a Stokes line. We will discus the Stokes phenomena in great detail in the Appendix \ref{Stokes}. Note that the contour \(\mathcal{J}_1\) can be continuously deformed into the line segment \([0,1]\), so evaluating the integral on \(\mathcal{J}_1\) produces the Virasoro vacuum block. However, understanding \(\mathcal{J}_2\) is a non-trivial task. It \(\mathcal{J}_2\) contour can be interpreted as the linear combination of the vacuum block and the \(\mathcal{O}_{3,1}\) block. Let's consider the integral along the contour \([\frac{1}{z},\infty)\) \cite{DiFrancesco:1997nk},
\begin{align}
    \mathcal{I}_{\pm}=& \oint dw \ w^a (1-w)^b (w-z)^c \nonumber \\ &\propto \ \mathcal{I}(-\infty,0)+ e^{\pm i\pi a}\mathcal{I}(0,1)+e^{\pm i\pi (a+b)}\mathcal{I}(1,\frac{1}{z})+e^{\pm i\pi (a+b+c)}\mathcal{I}(\frac{1}{z},\infty) \label{integral}
\end{align}
\begin{figure}
    \centering
    \includegraphics[width=0.60\linewidth]{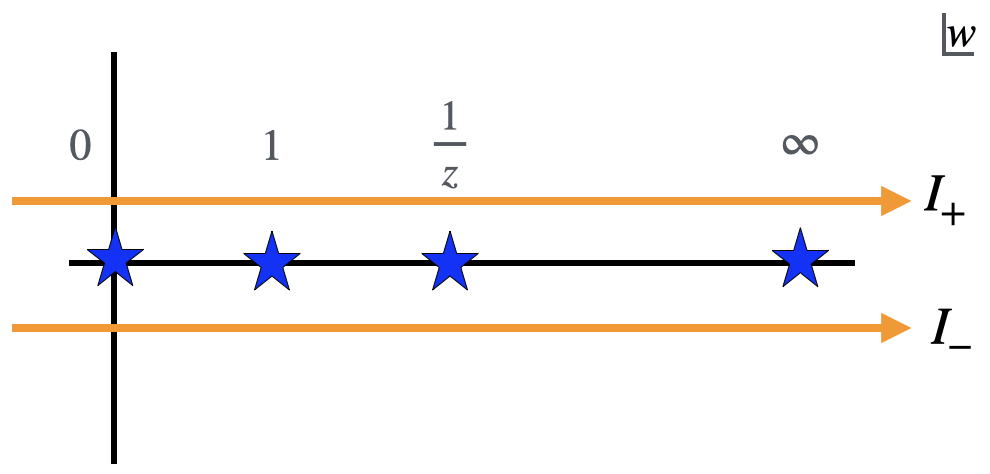}
    \caption{The figure displays the contours of integration \(I_{\pm}\) for \eqref{integral}. We will evaluate \(\mathcal{I}_+ - \mathcal{I}_-=0\) to establish a connection between the thimbles \(\mathcal{J}_i\) to Coulomb Gas integral over \([\frac{1}{z},\infty)\). }
    \label{fig:contour}
\end{figure}
When \(a+b+c<-1\), we can close the integral contours \(\mathcal{I}_{\pm}\) at infinity to obtain \(\mathcal{I}_+ - \mathcal{I}_- =0\) and relate the contour \([\frac{1}{z},\infty)\) to a contour encompassing the points \(0,1,\) and \(\frac{1}{z}\) as depicted in Figure \ref{fig:contour}. Working in the large \(c \propto b^2\) limit and approaching the forbidden singularity at \(z=2\) (i.e \(\theta=\pi\)), the branch point \(w=\frac{1}{z}\) and the steepest contours \(\mathcal{J}_1\) and  \(\mathcal{J}_2\) get closer to the real \(w\) axis. The vacuum block integral diverges when,
\begin{equation}
    b \rightarrow \infty \;\ ,\;\ \theta \rightarrow \pi
\end{equation}
The action \(\mathcal{I}_{2,1}\) \eqref{I} has a singularity at \(w=\frac{1}{z}\), so the integration contour must be deformed around this singularity. The difference between the contour integral on the upper half plane and the lower half plane in the interval \([\frac{1}{z},\infty)\) is equal to the sum over the deformed integral contours \(\mathcal{J}_1 - \mathcal{J}_2\) that encloses the remaining branch points \(0,1,\text{and} \frac{1}{z}\). Thus our contour integral becomes,
\begin{equation}
    \int_{\mathcal{J}_1 - \mathcal{J}_2} dw \ {\mathcal{I}_{2,1}} \propto \frac{(z-2)(z-1)^{b^2 -1}}{z^{2b^2+1}}
\end{equation}
Every time we cross a forbidden singularity, we cross a Stokes line (discussed in detail in Appendix \ref{Stokes}), shifting the vacuum Virasoro block to a heavy state \(\mathcal{O}_{3,1}\) Virasoro block. Alternatively, the contour integral over \(\mathcal{J}_1 - \mathcal{J}_2\) can be interpreted as a one \enquote{instanton} correction to the perturbative series of the vacuum conformal block.
\subsection{Heavy-light Blocks with \(\mathcal{O}_{3,1}\)}
In this section, we will deal with another example of a heavy-light Virasoro block with external operators \(\mathcal{O}_{3,1}\). The fusion rule gives us,
\begin{equation}
    \mathcal{O}_{3,1} \times \mathcal{O}_{3,1} = \mathbb{1}+\mathcal{O}_{3,1}+\mathcal{O}_{5,1}
\end{equation}
So the critical points in the Coulomb integral will be connection to two distinct heavy states. We can write the four-point function as,
\begin{equation}
     \braket{\mathcal{O}_{3,1}(\infty)\mathcal{O}_L(0)\mathcal{O}_L(z)\mathcal{O}_{3,1}(1)} \sim \int dw_1 \int dw_2 \frac{(1-z)^2}{z^{2h_{3,1}}} e^{\mathcal{I}_{3,1}}
\end{equation}
again for simplicity, we take \(h_L=1\) for which the action takes the form,
\begin{equation}
    \mathcal{I}_{3,1}(w) \equiv b^2\log[w_1(1-w_1)w_2(1-w_2)]-2b^2\log(w_1 -w_2)-2\log[(1-zw_1)(1-zw_2)]
\end{equation}
We will once again have singularities at \(w_i=\frac{1}{z_i}\), which will force us to deform the contours as we analytically continue in \(z\). Notice, that the action \(\mathcal{I}_{3,1}\) is symmetric under \(w_1 \leftrightarrow w_2\) in the complex plane. There are three different contour combinations for this Coulomb integral associated to this heavy-light block, namely \([0,1] \times [0,1]\), \([0,1] \times [\frac{1}{z},\infty)\) and \([\frac{1}{z},\infty) \times [\frac{1}{z},\infty)\). We will work in the OPE limit of \(z\) (i.e. \(z \rightarrow 0\)) in order to write these integrals in terms of the heavy operators found in the fusion rules,
\begin{equation}
    z^{-2h_{3,1}}(1-z)^2 \int_{0}^{1} dw_1 \int_{0}^{1} dw_2 \ e^{\mathcal{I}_{3,1}} \propto z^{-2h_{3,1}}
\end{equation}
So the contour choice \([0,1] \times [0,1]\) yields the Virasoro vacuum block. By a similar analysis, we conclude that the remaining two contour choices \( [0,1] \times [\frac{1}{z},\infty)\) and \([\frac{1}{z},\infty) \times [\frac{1}{z},\infty)\) will produce the heavy operators \(\mathcal{O}_{3,1}\) and \(\mathcal{O}_{5,1}\) respectively. We move onto understanding which critical points and heavy states resolves the forbidden singularities. Earlier, in Section \ref{borel}, we saw that the Borel resummation became ill-defined near the forbidden singularity due to a branch cut in the Borel plane. We want a similar analysis using the Coulomb gas formalism. Thus, we need to understand what happens to the integration contours when we analytically continue \(z\) towards the forbidden singularity. \\ Let the Lefschetz thimble be \(\mathcal{J}^{(j)}_i\) with \((j=1,2)\) and the associated critical points \(p_i\). There are six critical points, however due to the symmetry of the action \(w_1 \leftrightarrow w_2\), the critical points can be clubbed in three pairs. The critical points \(w_1,w_2\) are located at \(p_i=\left(\frac{1}{2}+ia_i, \frac{1}{2}+ib_i\right)\) with \(i=1,2,3\), where \(a_i,b_i\) are explicitly solved for in \cite{fitzpatrick2016informationlossads3cft2}.
\begin{figure}[t]
    \centering
    \includegraphics[width=0.80\linewidth]{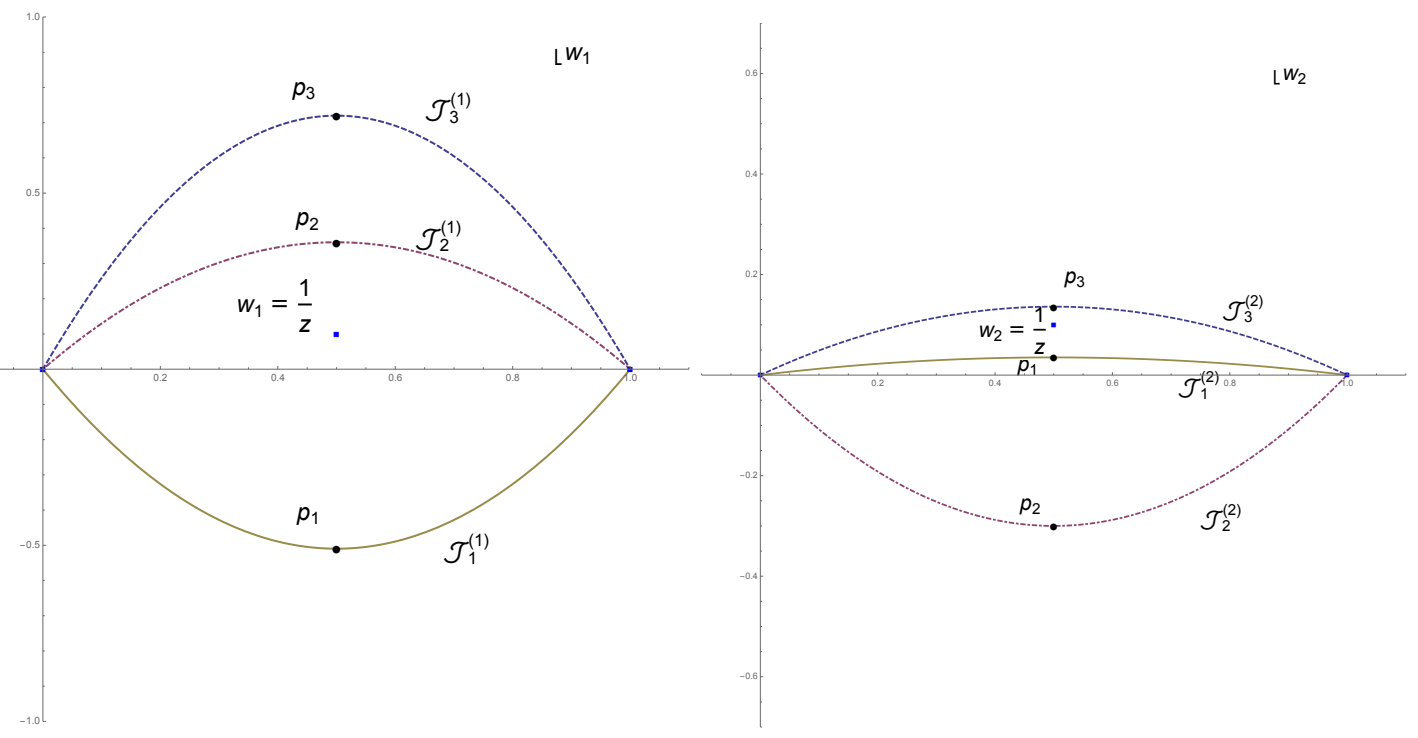}
    \caption{This figure illustrates the critical points (black dots) of the action \(\mathcal{I}_{3,1}\) and the steepest descent contours are defined using coloured curves. Here we have chosen \(b^2=10\). The contour \(\mathcal{J}^{(j)}_1\) (yellow curve) can be continuously deformed into \([0,1] \times [0,1]\) without passing through branch points \(w_i=\frac{1}{z_i}\) (blue dot), hence it corresponds to the vacuum block \cite{fitzpatrick2016informationlossads3cft2}.}
    \label{fig:enter-label}
\end{figure}
For the contour choice \( [0,1] \times [\frac{1}{z},\infty) \), the steepest contour \(\mathcal{J}^{(2)}_1\) (the yellow curve in \(w_2\) complex plane) needs to be deformed around \(\frac{1}{z}\), developing a heavy operator \(\mathcal{O}_{3,1}\).
\subsection{Heavy-light Blocks with \(\mathcal{O}_{2,2}\)}
In this section, we will deal with the final example of a heavy-light Virasoro block with external operators \(\mathcal{O}_{2,1}\). The fusion rule gives us,
\begin{equation}
    \mathcal{O}_{2,2} \times \mathcal{O}_{2,2} = \mathbb{1}+\mathcal{O}_{3,1}+\mathcal{O}_{1,3}+\mathcal{O}_{3,3}
\end{equation}
The fusion produces a light operator \(\mathcal{O}_{1,3}\) and two heavy operators \(\mathcal{O}_{3,1}\) and \(\mathcal{O}_{3,3}\). In the large \(c\) limit, we can approximate \(h_{2,1} \sim h_{2,2}\), so in this case we will have a single forbidden singularity at \(z=2\). Using the Coulomb integral we can write the four-point function as,
\begin{equation}
     \braket{\mathcal{O}_{2,2}(\infty)\mathcal{O}_L(0)\mathcal{O}_L(z)\mathcal{O}_{2,2}(1)} \sim \int dw_1 \int dw_2 \frac{(1-z)^{1+\frac{1}{b^2}}}{z^{2h_{2,2}}} e^{\mathcal{I}_{2,2}}
\end{equation}
again for simplicity, we take \(h_L=1\) for which the action takes the form,
\begin{align}
    \mathcal{I}_{2,2}(w) &\equiv (1+b^2)\log[w_1(1-w_1)]+(1+\frac{1}{b^2})\log[w_2(1-w_2)] \nonumber \\ &-2\log(w_1 -w_2)-2\log[(1-zw_1)]-\frac{2}{b^2}\log[(1-zw_2)]
\end{align}
\begin{figure}[h]
    \centering
    \includegraphics[width=0.83\linewidth]{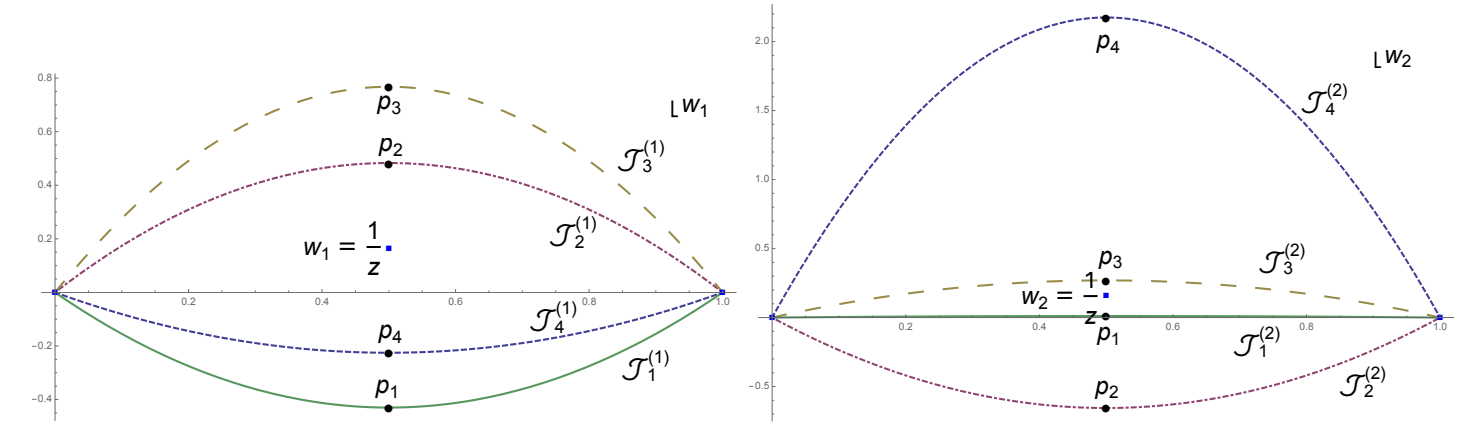}
    \caption{This figure illustrates the critical points (black dots) of the action \(\mathcal{I}_{2,2}\) and the steepest descent contours are defined using coloured curves. Here we have chosen \(b^2=3\). The contour \(\mathcal{J}^{(j)}_1\) (green curve) can be deformed into \([0,1] \times [0,1]\) without passing through branch points \(w_i=\frac{1}{z_i}\) (blue dot), so it corresponds to the vacuum block. As \(z\) revolves around 1 it induces a monodromy for the vacuum block \cite{fitzpatrick2016informationlossads3cft2}.}
    \label{fig:enter-label}
\end{figure}
The action has singularities at \(w_i=\frac{1}{z_i}\) which forces us to deform the contours as we analytically continue in \(z\). Since the action \(\mathcal{I}_{2,2}\) is not symmetric under \(w_1 \leftrightarrow w_2\) in the complex plane, we will have four different contour combinations namely \([0,1] \times [0,1]\), \([\frac{1}{z},\infty) \times [0,1]\), \([0,1] \times [\frac{1}{z},\infty)\) and \([\frac{1}{z},\infty) \times [\frac{1}{z},\infty)\). Once again by working in the small \(z\) limit, we can write,
\begin{equation}
   z^{-2h_{2,2}} (1-z)^{1+\frac{1}{b^2}} \int_{0}^{1} dw_1 \int_{0}^{1} dw_2 \ e^{\mathcal{I}_{2,2}} \propto z^{-2h_{2,2}}
\end{equation}
corresponds to the Virasoro vacuum block, while the contours \([\frac{1}{z},\infty) \times [0,1]\), \([0,1] \times [\frac{1}{z},\infty)\) and \([\frac{1}{z},\infty) \times [\frac{1}{z},\infty)\) correspond to \(\mathcal{O}_{3,1}\), \(\mathcal{O}_{1,3}\) and \(\mathcal{O}_{3,3}\) respectively. For the \(i\)th critical point in the \(w_j\) variable, we denote the Lefschetz thimbles as \(\mathcal{J}^{(j)}_i\) with \((j=1,2)\) and the associated critical points \(p_i\). In the parametrisation \(z=1-e^{i\theta}\), the coordinates of the four critical points is given by \(p_i=\left(\frac{1}{2}+ic_i,\frac{1}{2}+id_i \right)\), which are explicitly calculated in \cite{fitzpatrick2016informationlossads3cft2}. \\
Once again, we can continuously deform the steepest descent contours \(\mathcal{J}^{(j)}_1\) into \([0,1] \times [0,1]\) without passing through the forbidden singularity. So the integral over the Lefschetz thimble \(\mathcal{J}_1\) corresponds to Virasoro vacuum block. When the contours intersect \(\mathcal{J}_1\), the Coulomb integral becomes ill-defined, so we need to deform the contour to avoid the \(w=\frac{1}{z}\) point. In the present case, we must deform the contours in both \(w_1\) and \(w_2\) plane, picking a linear combination of states of the fusion \(\mathcal{O}_{2,2} \times \mathcal{O}_{2,2}\).
\chapter{Summary}
Let us briefly summarise the main results of the chapters and discussion the possible future research directions. We have shown that through CFT bootstrap for 2d large central charge theories we are able to involve some non-perturbative effects which match the dynamics of the dual AdS\(_3\). By the exchange of the identity block and subsequently its descendants, we obtain the dual multi-graviton intermediate states. This seems to the equivalent to presence of a deficit angle or a BTZ black hole in the background AdS\(_3\) geometry. A class of primaries with conformal dimension \(h_{\mathcal{O}}>c/24\), defined as heavy operators play a crucial role in making the AdS geometry to back react in which the light operators having dimension \(h_{\mathcal{O}}\ll c\) produce thermal correlator functions. Often in the community, this property is associated to the proof of the Eigenstate Thermalization Hypothesis for 2d CFT's in the semi-classical large \(c\) limit. Hitherto these corrections have been related to the eigenstate thermalization and more commonly the black hole “no hair”. \\ \\ Moreover, there is an interesting situation where we consider the light operators to be the “twist” operators sitting at the edge of an interval. In this case, one might be tempted to do a full bulk reconstruction in the heavy operator background using the Ryu-Takayanagi (RT) formula. However, there is a subtlety in such computations. We know that loop corrections to the Renyi entropy contaminate even a purely gravitational theory. These loop correction terms are a result of taking into account the effects of not only the vacuum block but also the contributions of it's Virasoro descendants. \\ \\ In the final chapters we focus on Information loss and it's resolution. Information loss in AdS black hole backgrounds infects correlators of the CFT with a set of sharply defined singularities. The two kinds we have discussed here are: forbidden singularities in Euclidean CFT correlators \cite{fitzpatrick2016conformalblockssemiclassicallimit}, and exponential decay behaviour in the late Lorentzian time regime \cite{Maldacena_2003}. These forbidden singularities that occur in the large \(c \propto \frac{R_{AdS}}{G_N}\) limit induces information loss effects in AdS\(_3\)/CFT\(_2\). However, we have shown that using degenerate conformal operators can lead to the prevention of information loss. Consequently, this helps get rid of the forbidden singularities of the Virasoro blocks via \(1/c\) type non-perturbative effects. It is expected that the Virasoro blocks can be expanded perturbatively in \(1/c\) using the gravitational or Chern-Simons path integrals \cite{hijano2015worldlineapproachsemiclassicalconformal,witten1998antisitterspaceholography,Witten:1988hf,Gaiotto_2012}. Essentially, this connection implies means that with a large central charge \(c\) and fixed ratios \(\eta \equiv h/c\), the perturbative series large \(c\) limit must be of the form,
\begin{equation}
    \mathcal{V}(h_i,\eta_I,c;z) \approx e^{-\frac{c}{6}f(\eta_i,\eta_I;z)}\sum_{n=0}^{\infty}\frac{1}{c^n}g_n(\eta_i,\eta_I;z)
\end{equation}
where the function \(f\) be finite in the semi-classical \(c \rightarrow \infty\) limit and \(h_I\) represent the intermediate primary states. \\ \\  Another intriguing connection we highlighted here was showing that singularities generated by the Borel series \(\mathcal{B}(s)\) due to the Stokes phenomena \cite{witten2010analyticcontinuationchernsimonstheory} in the Borel plane are  related to the semiclassical saddles. It is interesting to note that the semiclassical saddles are related to the Einstein's equations of motion. Analytically continuing away from this limit cross is analogous to crossing Stokes lines and thereby taking contributions from the sub-leading saddles. By taking into account the non-perturbative effects into the \(1/c\) perturbative expansion, the Virasoro block \(\mathcal{V}\) as a \enquote{transseries} \cite{Dorigoni_2019} becomes
\begin{equation}
    \mathcal{V}(h_i,\eta_I,c;z) = \sum_{p=0}^{\infty}\sum_{n=0}^{\infty} e^{-\frac{c}{6}f(\eta_i,\eta_I;z)}\frac{g_{p,n}(\eta_i;z)}{c^n}
\end{equation}
with the interpretation that the sub-leading terms \(p>0\) are the instanton contributions in AdS\(_3\). \\ \\ Finally, we investigate the Coulomb gas formalism. The advantage of using the vertex operator formalism because we expect there to be a strong connection between the Coulomb Gas with degenerate operators and the Chern-Simons path integral. Despite the correspondence between the physics of strings and black holes, if information loss in AdS\(_3\)/CFT\(_2\) strictly only depends on semi-classical blocks, string theory will play no role in the information restoration. 
\clearpage
\appendix
\chapter{Stokes Phenomena} \label{Stokes}
In the appendix, we will discuss the Stokes phenomenon by using Airy function as an example. For real arguments \(z\), the Airy function can be defined by,
\begin{equation}
    Ai(z)=\frac{1}{2\pi}\int_{-\infty}^{\infty} dt \ e^{\frac{i}{3}t^3+izt}
\end{equation}
The Airy's integral takes the same form with we have a complex \(z\) but with a modified contour. We will need to study the asymptotic behaviour of the Airy function to understand the Stokes phenomenon. \\ We use the method of stationary phase to the integral to define the Airy integral,
\begin{equation}
    Ai(z)=\frac{1}{2\pi}\int_{-\infty}^{\infty} dt \ e^{iS(z,t)}
\end{equation}
where \(S(z,t)\equiv \frac{1}{3}t^3+zt\) denotes the phase. We expand the phase, 
\begin{equation}
    S(z,t) \approx S(z,t=t_s)+\frac{1}{2}\frac{\partial^2 S}{\partial t^2} \Bigg|_{(t=t_s)} (t-t_s)^2
\end{equation}
around the points,
\begin{equation}
    t_s=\pm \sqrt{-z} \label{t_s}
\end{equation}
and the condition for the stationary phase,
\begin{equation}
    \frac{\partial S}{\partial t}\Bigg|_{(t=t_s)}=t^2_s +z=0
\end{equation}
We will confine ourselves to the real values \(z=x\). Using \eqref{t_s}, for \(x>0\) the points of the stationary phase \(t_s=\pm i\sqrt{|x|}\) are pure imaginary. In contrast to when \(x<0\), the stationary phase points \(t_s=\pm \sqrt{|x|}\) are purely real. When \(x=0\), the two asymptotic solutions of the stationary phase collapse into one. \\ \\ \textbf{Oscillatory region}: For the case when \(x<0\), we obtain an oscillatory solution to the Airy integral. Without providing a complete proof, we will just quote the results for our purpose,
\begin{equation}
    Ai(x) \approx \frac{1}{\sqrt{\pi}}|x|^{-\frac{1}{4}}\cos{ \left(\frac{2}{3}|x|^{\frac{3}{2}}-\frac{\pi}{4}\right)}
\end{equation}
\textbf{Decaying region}: For the case when \(x>0\), we obtain an exponentially decaying solution to the Airy integral,
\begin{equation}
    Ai(x) \approx \frac{1}{\sqrt{\pi}}|x|^{-\frac{1}{4}} e^{-\frac{2}{3}|x|^{\frac{3}{2}}}
\end{equation}
\textbf{Stokes phenomena}: Suppose \(g(z)\) is the asymptotic representation of the function \(f(z)\) as \(z \rightarrow z_0\). Then \(f(z)-g(z)\) is order \(O(g(z))\) in the asymptotic limit, thus \(f(z)=g(z)+(f(z)-g(z))\) \cite{john_heading_1962}.  When this condition is satisfied, it is said that when \(z\) lies in a certain sector where \(g(z)\) is dominant and \(f(z)-g(z)\) is sub-dominant. As \(z\) approaches the boundary of this sector, the behaviour flips between the dominant and the recessive functions. This effect is what's known as Stokes phenomenon and the edges of the sector are known as Stokes lines. \\ Suppose we are considering two independent solutions \(f_1\) and \(f_2\) of a 2nd order differential equation which have the leading behaviour as \(e^{S(z_1)}\) and \(e^{S(z_2)}\) respectively. Moreover, we will be assuming that the behaviour of the two solutions is quite different in different domains of the complex plane. There are domains where one solutions is more dominant than the other. The Stokes and Anti-Stokes lines define these domains. \\ Anti-Stokes lines in a complex plane are defined by,
\begin{equation}
    \mathrm{Re}[S_1 (z)-S_2 (z)]=0
\end{equation}
Hence, on an Anti-Stokes line, \(f_1\) and \(f_2\) have the same magnitude. \\ In contrast a Stokes lines is defined by,
\begin{equation}
     \mathrm{Im}[S_1 (z)-S_2 (z)]=0
\end{equation}
Stokes line clearly differentiates between the behaviour of the two solutions as,
\begin{equation}
    f_1 \sim \exp({\mathrm{Re}S_1 (z)})e^{i\phi} \;\ , \;\  f_2 \sim \exp({\mathrm{Re}S_2 (z)})e^{i\phi}
\end{equation}
and \(\phi=\mathrm{Im}S_1(z)=\mathrm{Im}S_2 (z)\). When we cross a Stokes line the two solutions exchange their character. They go from a dominant (sub-dominant) to a sub-dominant (dominant) solution. 
\begin{figure}
    \centering
    \includegraphics[width=0.35\linewidth]{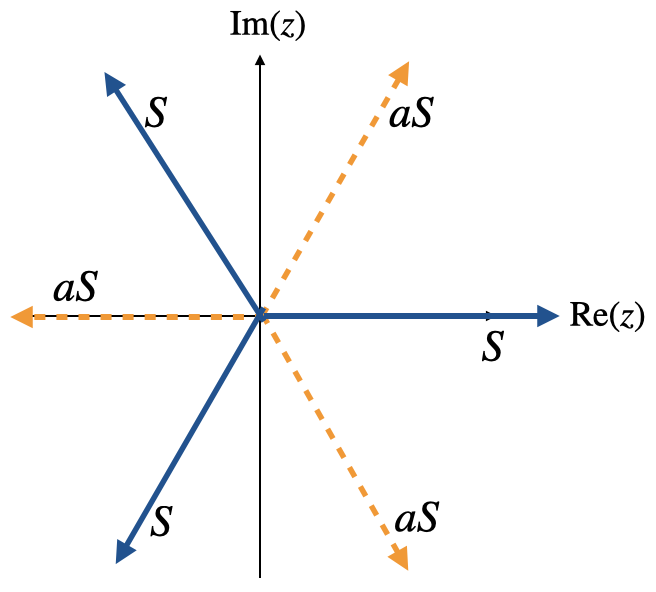}
    \caption{Stokes (S) and Anti-Stokes (aS) lines of the Airy Function in a complex plane.}
    \label{fig:sasfig}
\end{figure} 
\\ \textbf{Application to Airy function}: The differential equation for the Airy function is,
\begin{equation}
    f^{''}-zf=0
\end{equation}
and has two independent solutions \(f_1 =Ai(z)\) and \(f_2 = Bi(z)\). We have shown that the asymptotic behaviour of the solutions are given by,
\begin{align}
    &f_1 \sim z^{-1/4} \exp \left({-\frac{2}{3}z^{3/2}}\right) \nonumber \\ &f_2 \sim z^{-1/4} \exp \left({\frac{2}{3}z^{3/2}}\right)
\end{align}
Hence the condition \(\mathrm{Re}(z^{3/2})=0\) defines the Anti-Stokes lines,
\begin{equation}
    \arg{z}=\pm \frac{\pi}{3},\pi
\end{equation}
While the condition \(\mathrm{Im}(z^{3/2})=0\) defines the Stokes lines,
\begin{equation}
    \arg{z}=0,\pm \frac{2\pi}{3}
\end{equation}
as shown in Figure \ref{fig:sasfig}. Hence the solution of the Airy function \(Ai(z)\) is sub-dominant in the sector \(\left(-\frac{\pi}{3}<\arg{z}<\frac{\pi}{3} \right)\) of the complex plane, depicted in Figure \ref{fig:sasfig}. Consequently the solution \(Bi(z)\) is dominant in the sector \(\left(-\frac{\pi}{3}<\arg{z}<\frac{\pi}{3} \right)\) ans sub-dominant in the complementary region.
\clearpage
\bibliographystyle{utphys}
\bibliography{ref}
\end{document}